\documentclass{aa}

\usepackage{graphicx}
\usepackage{rotating}
\usepackage{txfonts}
\usepackage{natbib}

\usepackage{bookmark}

\begin{document}

\title{Deep VISTA observations of the Magellanic Clouds (VMCDeep)}

\titlerunning{VMCDeep}

   	 \author{M.-R. L. Cioni \inst{1}
	 \and A. O. Omkumar \inst{1,2}     
         \and N. J. G., Cross \inst{3} 	 
         \and F. Cusano \inst{4}
         \and F. Dresbach \inst{5}	 	
         \and F. Ficara \inst{6,7}  
	 \and F. Niederhofer \inst{1}
          \and J. M. Oliveira \inst{5}   
          \and V. Ripepi \inst{7}
 	  \and S. Vijayasree \inst{1,2}                 
          \and G. Clementini \inst{4}
          \and R. de Grijs \inst{8,9,10}
          \and L. Girardi \inst{11}
          \and M. A. T. Groenewegen \inst{12}
          \and V. D. Ivanov \inst{13}          
          \and N. Kacharov \inst{1}
          \and M. Marconi \inst{7}
          \and A. Mazzi \inst{14}
          \and G. Pastorelli \inst{15}
          \and C. M. Pennock \inst{3}
          \and M. Petr-Gotzens \inst{13}
          \and J. Storm \inst{1}
          \and S. Subramanian \inst{16}
          \and J. Th. van Loon \inst{5}   
	 \and R. P. Blake\inst{3}
          \and M. J. Holliman\inst{3}
          \and J. Irwin \inst{17}
          \and M. J. Irwin \inst{17}
          \and R. G. Mann\inst{3}
          \and M. Read \inst{3}
          \and E. Sutorius\inst{3}
          }

   \institute{Leibniz-Institut f\"{u}r Astrophysik Potsdam, An der Sternwarte 16, D-14482 Potsdam, Germany\\
   	\email{mcioni@aip.de}
         \and Institut f\"{u}r Physik und Astronomie, Universit\"{a}t Potsdam, Haus 28, Karl-Liebknecht-Str. 24/25, D-14476 Potsdam, Germany 	
         \and Institute for Astronomy, Royal Observatory, Blackford Hill, Edinburgh EH9 3HJ, United Kingdom
         \and INAF -- Osservatorio di Astrofisica e Scienza dello Spazio, Via Gobetti 93/3, I-40129 Bologna, Italy
         \and Lennard-Jones Laboratories, Keele University, Keele ST5 5BG, United Kingdom
	\and Universit\`{a} di Salerno, Dipartimento di Fisica "E.R. Caianiello", Via Giovanni Paolo II 132, I-84084 Fisciano (SA), Italy
	\and INAF -- Osservatorio Astronomico di Capodimonte, salita Moiariello 16, I-80131 Napoli, Italy
         \and School of Mathematical and Physical Sciences, Macquarie University, Balaclava Road, Sydney, NSW 2109, Australia
	\and Astrophysics and Space Technologies Research Centre, Macquarie University, Balaclava Road, Sydney, NSW 2109, Australia
	\and International Space Science Institute--Beijing, 1 Nanertiao, Zhongguancun, Hai Dian District, Beijing 100190, China   
	\and INAF -- Osservatorio Astronomico di Padova, Vicolo dell'Osservatorio 5, I-35122 Padova, Italy
         \and Koninklijke Sterrenwacht van Belgi\"{e}, Ringlaan 3, B-1180 Brussels, Belgium
	\and European Southern Observatory, Karl-Schwarzschild-Str. 2, D-85748 Garching bei M\"{u}nchen, Germany   	      
	\and Dipartimento di Fisica ed Astronomia Augusto Righi, Universit\`{a} degli Studi di Bologna, via Gobetti 93/2, I-40129 Bologna, Italy	   
	\and Astronomical Observatory, University of Warsaw, Al. Ujazdowskie 4, 00-478 Warsaw, Poland
         \and Indian Institute for Astrophysics, II Block, Koramangala, Bengaluru 560 034, India		
         \and Institute of Astronomy, University of Cambridge, Madingley Road, Cambridge CB3 0HA, United Kindgom      
             }
             	
\date{Received February 22, 2026; accepted August 27, 2026}

\abstract
  {The central regions of galaxies challenge observers with a high density of stars, gas and dust. The Large Magellanic Cloud (LMC) hosts a dense stellar bar whereas the Small Magellanic Cloud (SMC) hosts a dense core, which have eluded even the most sensitive large-scale wide-field surveys of individual stars from the ground and from space.}
  {This work describes deep imaging observations obtained with the Visual and Infrared Survey Telescope for Astronomy (VISTA) of the inner regions of the Clouds. They were obtained to resolve stars fainter than the red clump, that due to crowding were missed in previous observations, and to quantify features associated with the star formation, kinematics and structure of the galaxies. These studies will allow us to establish how the dynamical interactions between the LMC and the SMC, and the interaction between the Clouds and the Milky Way have influenced the most gravitationally bound regions.}
  {Multiple near-infrared images were obtained in the $J$ and $K_\mathrm{s}$ filters during excellent weather conditions (e.g.\,with a seeing of $\sim$0.7 arcsec). The data were reduced using the VISTA Data Flow System which provides single-filter and band-merged calibrated images and catalogues, with associated confidence maps, for individual observations and combined observations for deep tiles.}
  {Approximately one million and 920 thousand sources, with only minor quality issues (i.e.\,by excluding sources with a problematic photometric calibration or location on the detectors) and about 40 epochs per filter, are detected in the central 1.77 deg$^2$ region of the LMC and SMC, respectively. They provide a more contiguous spatial area coverage than sources in the VISTA survey of the Magellanic Clouds system (VMC), which covers the same regions by overlapping two VISTA tiles per galaxy. Two thirds of these sources are likely stars whereas the other 1/3 could be stellar blends or background galaxies. The catalogues, which are made available with this work, contain attributes to select unique good quality detections for which we provide examples of ongoing scientific applications.}
   {These data represent valuable complementary information to deep imaging observations at other wavelengths and provide additional epochs for sources in common with the VMC survey, for proper motion and photometric variability studies extending the time baseline to $\sim$11 yr, as well as targets for spectroscopic observations.}
   

\maketitle
\nolinenumbers

\section{Introduction}
\label{introduction}

The inner parts of galaxies are among the most challenges places to observe individual stars because they are prone to crowding, contain gas and dust and may host nuclear structures such as a black hole. The Large and Small Magellanic Clouds (LMC and SMC) are our nearest pair (50$-$60 kpc distance; \citealp{degrijs2014} and \citealp{degrijs2015}) of interacting dwarf irregular galaxies and locating their gravitational centres is not trivial due to their mutual dynamical interaction and their interaction with the Milky Way (e.g. \citealp{bekki2008}, \citealp{besla2012}, \citealp{rathore2025b,rathore2026}). These events may cause the Clouds to break apart (e.g.\,\citealp{deason2017}, \citealp{2018NatAs...2..901M}) and we are witnessing an early stage of this process. The Clouds are on their first (e.g. \citealp{kallivayalil2013}, \citealp{conroy2021}) or perhaps second (\citealp{patel2017}, \citealp{vasiliev2024}) passage by the Milky Way.

Several multi-wavelength imaging surveys have produced data to unravel properties of the Clouds, their star formation history (SFH), three-dimensional (3D) structure, rotational pattern, tidal tails, and associated satellites (e.g.\,\citealp{sales2013}, \citealp{ripepi2017}, \citealp{helmi2018}, \citealp{mackey2018}, \citealp{2018MNRAS.478.5017R}, \citealp{kallivayalil2018}, \citealp{belokurov2019}, \citealp{elyoussoufi2019}, \citealp{2021MNRAS.tmp.2166M}, \citealp{massana2022}, \citealp{cullinane2023}). 
Despite major progress in characterising their external regions, our observational capabilities deteriorate in the central regions. The most sensitive ground-based imaging survey to-date, the Survey of the Magellanic Clouds Stellar History (SMASH; \citealp{2017AJ....154..199N}), shows a lack of stars in the LMC centre \citep{2018ApJ...866...90C} due to crowding.  The most powerful space astrometry mission, {\it Gaia}, is also incomplete in the high-density regions \citep{prusti2016} although there are in general $\sim$20\% more sources in Data Release 3 (DR3) than in DR2 \citep{vallenari2023}. For example, the incompleteness within a 2.5 deg radius of the LMC centre is at least 50\% (\citealp{jimenez-arranz2023}, \citealp{rathore2025a}).
The most extensive monitoring programme (in the V and I bands), the Optical Gravitational Lensing Experiment (OGLE; \citealp{2015AcA....65....1U}), shows that an apparent extension along the line of sight, as traced by RR Lyrae stars in the inner region, is the result of blending effects \citep{Jacyszyn-Dobrzeniecka2017,Jacyzyn-Dobrzeniecka2020}. The Visual and Infrared Survey Telescope for Astronomy (VISTA; \citealp{sutherland2015}) survey of the Magellanic Clouds system (VMC; \citealp{cioni2011}), because of its sensitivity and spatial resolution in the near-infrared, allows to penetrate the dusty and dense regions, but not adequately enough (i.e.\,sources are still blended and faint stars are not detected) to quantify structures related to the origin, dynamics, and evolution of the galaxies \citep{niederhofer2022}. Even the central parts of the Clouds are very extended on the sky and accurate studies with the {\it Hubble Space Telescope} (HST), (e.g.\,\citealp{kallivayalil2013}, \citealp{weisz2013}, \citealp{vandermarel2014}, \citealp{zivick2018}, \citealp{roman2019}, \citealp{murray2024}), or the {\it James Webb Space Telescope} (JWST) over several square degrees are impractical, given their small field of view (HST$\sim7$ arcmin$^2$, JWST$\sim5$ arcmin$^2$).

The location of the centre of the LMC, which depends on the type of tracer (stars or gas) and perhaps also on wavelength, is highly debated (e.g.\,\citealp{vandermarel2014}). The disc-centre may be displaced from the bar-centre due to the interaction with the SMC \citep{pardy2016} and there is only tentative evidence of the presence of a central black hole \citep{boyce2017,erkal2019,han2025}. {\it Gaia} Early DR3 (EDR3) data suggest that the rotational centre is closer to the centre from photometric studies than from the H\,{\sc i}  gas \citep{luri2021}. Using multi-epoch VMC data \cite{niederhofer2022} determined the proper motion of stars within the central parts of the LMC. They used a simple flat-rotating disc model to analyse and interpret the proper motion data and found a stellar centre of rotation very close ($\sim$10 arcmin) to the position of the tentative central black hole as well as a high inclination ($\sim$33.5 deg) of the central parts of the galaxy. Their study confirmed a higher rotational velocity for the young stars ($\sim$90 km s$^{-1}$ and <0.5 Gyr old) compared to the intermediate-age/old stars ($\sim$70 km s$^{-1}$ and >1 Gyr old), which can be explained by asymmetric drift. They constructed spatially resolved velocity maps of the different (by age) stellar populations which showed that intermediate-age/old stars follow elongated orbits parallel to the bar's major axis whereas young stars show motions along a central filamentary bar structure. 
In a subsequent study by \cite{vijayasree2025}, which extends the VMC time baseline from about six to ten years, the position of the centre was refined within the uncertainties of the previous determination. Their study, which covers the entire galaxy, following the analysis method of \cite{niederhofer2022}, allows to also derive a kinematic centre which turns out to be very close to the dynamical centre of mass obtained by \cite{rathore2025a}. These authors developed a novel solution to tackle crowding-induced incompleteness in the {\it Gaia} data in order to measure the bar properties (radius, axis ratio, position angle and strength) using red clump stars. They traced the crowding level using the {\it Gaia} BP--RP colour excess.

In the SMC \cite{niederhofer2018,niederhofer2021} showed that stars within the dense core region have a systematically different motion which can be explained either by stretching of the SMC, so the outer parts move differently than the central parts, or by a tidal feature behind the main body of the galaxy. Data from HST support SMC stretching \citep{zivick2018} and the {\it Gaia} data confirm a proper motion dichotomy \citep{luri2021}. The apparent anisotropic tangential motion at all radii derived from {\it Gaia} DR2 data and spectra of many giant stars corroborate the tidal disruption of the SMC, possibly all the way into the core \citep{deleo2020}. The SMC internal structure and kinematics is largely explained by a recent ($\sim$100 Myr ago) LMC-SMC collision \citep{rathore2026}.

Motivated by the excellent astrometric performance of the VISTA camera and inspired by the possibility of finding the cores of the Clouds, we developed an observational programme to obtain more robust proper motions to constrain the central kinematic patterns. We focused on one VISTA {\it tile} (a region of $\sim$1.77 deg$^2$) per galaxy and were awarded observations at the European Southern Observatory (ESO) with a better sky quality and an increased  number of epochs, at both $J$ and $K_\mathrm{s}$ bands,  with respect to the VMC survey. These aspects, including a long time-baseline obtained from the combination of deep and shallow (from VMC) data, will provide an independent and more precise measurement of the stellar proper motions in the densest regions of the Clouds. 
In this paper, Section \ref{observations} describes the observations and Section \ref{reduction} presents the data reduction process. A comparison between the new data and those from the VMC survey is given in Section \ref{difference}. Section \ref{quality} presents the content of the catalogue whereas ongoing scientific applications of the data are presented in Section \ref{content}. Section \ref{summary} concludes this study.

\section{Observations}
\label{observations}

\begin{figure*}
\centering 
\includegraphics[width=20cm,height=10cm]{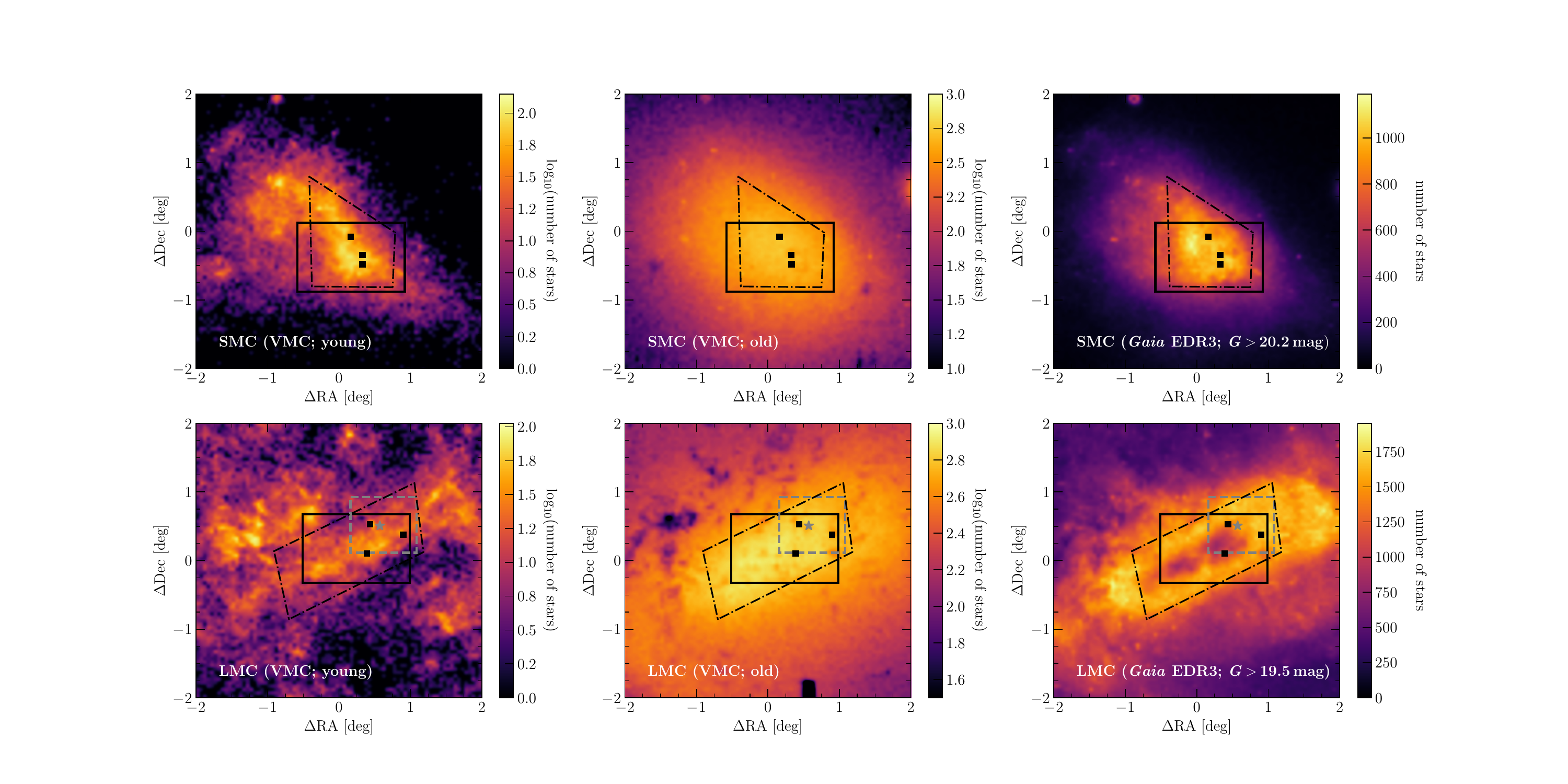}
\caption{Distribution of young (upper main sequence stars, left) and old (red clump and brighter stars, middle) stars from the VMC survey and {\it Gaia} EDR3 (right) in $3\times3$ arcmin$^2$ bins across the SMC (top) and LMC (bottom). The solid rectangles indicate the VISTA tiles from this programme while the irregular polygons show the areas with special {\it Gaia} observations designed to overcome crowding issues (see text for details). Small black squares highlight star forming regions while a dashed-grey box shows the location of a suspected central black hole (star).}
\label{map}
\end{figure*}

The VISTA data described here correspond to a total of 160 hours obtained during the period from December 2020 to January 2023 and refer to ESO programmes 105.206M, 106.201G, 108.221P and 109.23G7 (VMCDeep or this programme hereafter). VISTA is a 4 m class alt-azimuth mounting survey telescope which is part of ESO's Paranal Observatory. Until March 2023, VISTA was equipped with an infrared camera (VIRCAM; \citealp{sutherland2015}) operating between 0.8 $\mu$m and 2.3 $\mu$m with a set of broad and narrow band filters. VIRCAM had an array of 16 Raytheon detectors with a mean pixel size of 0.339 arcsec and a 1.65 deg diameter field-of-view. The point-spread function (PSF) of the instrument and camera systems was designed to have a full width at half maximum (FWHM) of 0.51 arcsec.
VISTA observed a continuous area of sky by filling in the gaps between the detectors using a sequence of offsets, each by a significant fraction of a detector. In this programme, we combine three fixed offset positions in the "Y" direction of the array, named {\it pawprints}, to cover an area of $\sim0.9$ deg$^2$ (a VISTA {\it half tile}); each pawprint covers an area of 0.6 deg$^2$ excluding the gaps between the detectors. Then, we combine another set of three pawprints which were offset also in the "X" direction, with respect to the previous set. Each half tile corresponds to a stripy coverage of the sky area which is counter-intuitive with respect to the nomenclature used to refer to them in the observations, i.e. left (LFT) and right (RHT). The merging of the two half tiles produces a VISTA {\it tile} that covers about 1.77 deg$^2$ (Fig.\,\ref{map}). 

Observations were obtained in service mode by ESO staff. This mode  guarantees efficiency of operations and a high level of data homogeneity, it is also the best method to capture images of high-quality which are needed for this study to disentangle sources in the dense stellar regions at the centres of the Clouds. We requested observations to take place during the following weather conditions: seeing of $0.7$ arcsec at $500$ nm at zenith, airmass $<1.7$, any lunar phase, and a variable/thin cirrus sky transparency. A violation of these observing requirements by $10$\% is still considered as if observations were obtained within specifications, whereas for larger deviations observations were repeated. Table \ref{table:observations} describes the main parameters of the VISTA observations which follow the nesting sequence FPJME (see VISTA user Manual\footnote{https://www.eso.org/sci/facilities/paranal/decommissioned/vircam/\-doc.html}). This sequence first sets a filter (F), then obtains images at all jittered positions of the first pawprint (P), before moving to the next pawprint and taking all the jittered images at that position (J), within a microstepping pattern (M) if specified and so on, until all multiple three pawprints that form the half tile are executed (E). The Jitter pattern was {\it Jitter9s} and the tile patterns were {\it Tile3nx} and {\it Tile3px} for the RHT and LFT half-tile observations, respectively. The $J$- and $K_\mathrm{s}$-band observations were independent from each other and could have been obtained during the same night without restrictions in time of execution except for the wish to distribute observations of the same half tile and filter across multiple nights. The integration time at a given Jitter position is calculated by multiplying the Detector Integration Time (DIT) by the number of DITs. Taking into account that each pawprint has 9 Jitter positions and each half tile has 3 pawprints, we obtain a total integration time per half tile of 540 s in $J$ and $1350$ s in $K_\mathrm{s}$.
Through the reconstruction of the tiling pattern most points of the sky are observed twice (on average). The exception are the tile edges, observed once, and some areas of extra overlap among the detectors that are observed four or six times. The total integration time, including observations at all sky conditions, was of 70 hours in the LMC and of 57 hours on the SMC with only about two hours lost due to technical problems that corrupted the images and resulted in incomplete half-tile observations, or produced images with FWHM>2 arcsec.

	\begin{table}
		\caption{Observing parameters.}                        
		\label{table:observations}      
		\small
		\begin{tabular}{lcc}
			\hline \hline
			Filter & $J$ & $K_\mathrm{s}$ \\
			\hline
		    Central wavelength ($\mu$m) & 1.25 & 2.15 \\
		    Bandwidth ($\mu$m) & 0.18 & 0.30 \\
		    Detector Integration Time -- DIT (s) & 10 & 5 \\
		    Number of DITs & 2 & 10 \\
		    Number of exposures & 1 & 1 \\
		    Microstepping & 1 & 1 \\
       	            Number of Jitters & 9 & 9 \\
                     Paw-prints in half tile & 3 & 3 \\
                     Pixel size (arcsec) & 0.34 & 0.34 \\
                     Integration time per pawprint (s) & 180 & 450 \\
                     Integration time per half tile (s) & 540 & 1350 \\
                     Number of LMC half tiles  & 149 & 127 \\
                     Number of SMC half tiles & 124 & 102\\
                     Total integration time in LMC (h) & 22 & 48 \\
                     Total integration time in SMC (h) & 19 &  38 \\
                     Saturation limit (Vega mag) & 12.7 & 11.4 \\
                     Tile area (deg$^2$)  & 1.77 & 1.77 \\
                     System FWHM (arcsec) & 0.51 & 0.51 \\
		\hline
		\end{tabular}
		\tablefoot{Jitter pattern = {\it Jitter9s}. Tile patterns = {\it Tile3nx} and {\it Tile3px}.}
	\end{table}

The pawprint mosaic was prepared using the survey area definition tool (SADT; \citealp{Arnaboldi2008}) using default values for the parameters (maximum jitter = 15 arcsec and {\it backtrackStep}=100) and also for the tile orientation (position angle = 0 deg) with the "Y" axis to the North and the "X" axis to the West. The position angle is defined to increase from North to East. This is the best orientation to maximise the coverage of the central areas of the Clouds where {\it Gaia} data are not optimal (Fig.~\ref{map}). We created two geodesic rectangles centred at $\alpha=05$:$20$:$54.47$, $\delta=-69$:$34$:$43.32$ (J2000.0) for the LMC and $\alpha=00$:$50$:$16.06$, $\delta=-73$:$10$:$40.92$ (J2000.0) for the SMC with width = 1.4753 deg and height = 1.2005 deg. Guide stars were assigned automatically to each tile using the Two Micron All Sky Survey (2MASS) catalogue \citep{skrutskie2006}.

\section{Data processing}
\label{reduction}

\subsection{CASU reduction}
The raw VISTA images were reduced using the VISTA Data Flow System (VDFS; \citealp{Irwin2004}) version 1.5 at the Cambridge Astronomy Survey Unit (CASU\footnote{http://casu.ast.cam.ac.uk/surveys-projects/vista}). The most relevant steps in the reduction of the VISTA images are: -- reset, dark, linearity, flat-field and de-stripe correction (i.e., removing a horizontal pattern in the background); -- sky-background correction; -- jitter and pawprint stacking; -- catalogue generation; -- astrometric and photometric calibration. Detector crosstalk and sky fringing as well as persistence effects following the observation of a bright object are negligible or rare and are ignored during pipeline processing. The reduction takes also into account information from all VISTA data obtained during the same night and week regardless of host programme. Observational uncertainties are propagated during the data processing and quality control parameters are calculated to monitor the data to both evaluate the observing conditions (in retrospect) and the individual data reduction steps. Among them are: the zero-point to measure the atmospheric extinction, the FWHM to measure the seeing, the ellipticity to evaluated the quality of the guiding and active optics correction, and the sky level to estimate the background level and its variations. Table \ref{table:data} shows the mean values of some quality control parameters for this programme. The quality parameters for observations that do not meet the sky quality conditions, and that were therefore repeated, are provided in Table \ref{table:data_bad}. These images are processed separately and are not stacked with those that instead satisfy the sky criteria.

Overall, 506 sets of pawprint images were obtained, where one set corresponds to the three offset positions that cover a given half tile. Four half tiles and six pawprints, that refer to incomplete half tiles, were discarded (see Sect.\,\ref{observations}). The others show an excellent consistency among the quality parameters associated to the observations of the two half tiles at the same wave band for both high (Table \ref{table:data}) and low (Table \ref{table:data_bad}) quality data.
A pawprint image is produced by stacking and combining the images obtained at each jitter position. Each image (including a confidence map) and catalogue are delivered in Rice compressed format and are multi-extension FITS files containing each the information for all of the 16 detectors covering the field-of-view. The processing history is recorded directly in the FITS headers. 

	\begin{table*}
		\caption{Quality control parameters for high-quality observations.}                        
		\label{table:data}      
		\small
		\begin{tabular}{lcccccccccc}
			\hline \hline
			Galaxy & $\alpha$ & $\delta$ & Filter & Half tile & N & Airmass & FWHM & Ellipticity & Zero-point & Sensitivity \\
			 & (h:m:s) & ($^\circ$:$^\prime$:$^{\prime\prime}$) & & & & & ($^{\prime\prime}$) & & (mag) & (mag) \\
			\hline
            SMC & 00:50:16.06 & --73:10:40.92 & $J$ & LFT & 43 & 1.54$\pm$0.04 & 0.80$\pm$0.06 & 0.08$\pm$0.02 & 23.80$\pm$0.11 & 19.86$\pm$0.13\\
             & & & $J$ & RHT & 42 & 1.57$\pm$0.09 & 0.81$\pm$0.07 & 0.08$\pm$0.02 & 23.82$\pm$0.07 & 19.89$\pm$0.14\\
             & & & $K_\mathrm{s}$ & LFT & 40 & 1.54$\pm$0.04 & 0.74$\pm$0.06 & 0.07$\pm$0.02 & 23.07$\pm$0.02 & 18.89$\pm$0.08\\
             & & & $K_\mathrm{s}$ & RHT & 40 & 1.56$\pm$0.05 & 0.74$\pm$0.05 & 0.07$\pm$0.02 & 23.07$\pm$0.02 & 18.88$\pm$0.07\\
            LMC & 05:20:54.47 & --69:34:43.32 & $J$ & LFT & 37 & 1.46$\pm$0.06 & 0.78$\pm$0.04 & 0.09$\pm$0.03 & 23.75$\pm$0.11 & 19.18$\pm$0.07 \\
             & & & $J$ & RHT & 38 & 1.48$\pm$0.07 & 0.79$\pm$0.04 & 0.09$\pm$0.02 & 23.75$\pm$0.09 & 19.21$\pm$0.06 \\
             & & & $K_\mathrm{s}$ & LFT & 38 & 1.47$\pm$0.06 & 0.76$\pm$0.05 & 0.07$\pm$0.02 & 23.03$\pm$0.03 & 18.46$\pm$0.06 \\
             & & & $K_\mathrm{s}$ & RHT & 38 & 1.47$\pm$0.07 & 0.75$\pm$0.05 & 0.07$\pm$0.02 & 23.02$\pm$0.03 & 18.49$\pm$0.05 \\
			\hline
		\end{tabular}
		\tablefoot{The coordinates are those of the tile centres. LFT corresponds to the tile pattern {\it Tile3px} and RHT to {\it Tile3nx}. N is the number of epochs. The sensitivity corresponds to 5$\sigma$ point source limiting magnitudes.}
	\end{table*}

\subsection{WFAU reduction}
The data produced at CASU using the VDFS are ingested into the VISTA Science Archive\footnote{http://vsa.roe.ac.ukl} (VSA; \citealp{cross2012}) which is hosted at the Wide Field Astronomy survey Unit (WFAU\footnote{https://ifa.roe.ac.uk/research-areas/wide-field-astronomy-unit}). At the WFAU the data are further processed and curated to produce standardised data products using also the VDFS, which guarantees that the data are processed homogeneously throughout the entire processing chain. Multiple epochs in the same filter and pointing are combined to produce {\it deepstack} images, which due to the increased exposure time allow to detect faint sources. A {\it tile} image in a given filter is created by combining all individual images in that filter (also from different pointings) that meet pre-defined quality criteria. Their sky level and individual pawprint astrometric and photometric distortions are adjusted in the grouting process \citep{carlos2018}. Deepstack and tile images have associated confidence images that contain the relative weighting for each pixel. This includes bad pixel masking, the effects of jittering, seeing and exposure time weighting. Further data products are generated in a database-driven manner. Catalogues are extracted from the deep images and single-filter catalogues are merged to create catalogues of sources, which can be cross-correlated with external catalogues (hosted at the VSA) to produce catalogues of neighbours.

The basic tables which contain information on the deep VISTA data are the {\it vmcdeepDetection}, which contains the catalogues corresponding to individual observations, and {\it vmcdeepSource}, which contains the list of sources obtained from deepstack images matched between the $J$ and $K_\mathrm{s}$ bands. In the vmcdeepSource table the same source may appear more than once depending on its location with respect to the overlap among the pawprints forming a tile. These duplicates are identified in a {\it seeming} process and the attribute {\it priOrSec} allows users to select unique sources. Note that the best detection among the duplicates is defined as the source closest to the optical axis of the camera. 
The position and magnitude of each source in any given table refers to the astrometrically and photometrically calibrated measurements using the parameters specified in the image headers. The photometric calibration of the VISTA data is described in \cite{carlos2018} and the precision achieved in the $J$ and $K_\mathrm{s}$ filters is better than 2\%. All tiles show a 10--20 mas systematic astrometric pattern. Both astrometry and photometry rely on the observation of stars from the 2MASS catalogue within each detector and with magnitudes between 12 and 14 in both bands. In addition, there are quality flags ({\it ppErrBits}) that identify problems related to the ingestion, incompleteness, and detection of the sources as well as morphological classification flags ({\it mergedClass}) to indicate stellar, non-stellar, borderline sources and noise. For more details about the columns present within each table readers are referred to the CASU and VSA web pages. Known issues present in the VISTA data, such as bad regions in specific detectors or spurious sources around bright stars, are also described on the web pages.  Only images that meet the observing requirements (Table \ref{table:data}) are included into the creation of deepstack and tile images as well as their associated catalogues. The other images (Table \ref{table:data_bad}) are deprecated and remain only available as single observations/detections. Figures \ref{lmcfig} and \ref{smcfig} show most of the LMC and SMC tiles produced from the stacking of all the good-quality observations.

\subsection{PSF photometry}
\label{sec:psf}

Point Spread Function (PSF) photometry is performed separately in the $J$ and $K_\mathrm{s}$ bands: per epoch, per half tile and on the stacked images, using IRAF/DAOPHOT tasks and following the method described in \cite{rubele2015}\footnote{The PSF is an independent processing of the data that does not use the output of the VDFS for source detection.}. The PSF processing was executed in a non-iterative mode after setting some parameters to values specific for the VMCDeep data and keeping the others as for the VMC survey, e.g.\,the PSF FWHM in pixels, the aperture radius and the PSF model radius were reduced according to Tab.\,\ref{table:quality}.
A  band-merged catalogue is subsequently created by associating the closest sources within 1 arcsec in distance. In brief the method works as follows. Individual pawprint images are first homogenised to account for variations of PSF across detectors. This is done by fitting a constant PSF model to a sample of bright and isolated non-saturated stars, which is then adjusted to the largest half-flux PSF radius. Each image is then convolved with a kernel that produces images corresponding to the (symmetric Moffat function) PSF model. The homogenised pawprint images at different epochs are subsequently combined using the SWarp programme \citep{bertin2002}, which resamples them onto a common astrometric grid and co-adds them using weights that account for the relative depth and image quality of each contributing exposure, to produce a uniform sky subtracted deep tile image.  Artefacts in the images are removed by masking contaminated regions. This process aims to render more uniform the limiting magnitude on the final deep tile with respect to intrinsic differences in detector sensitivity and stellar crowding. However, the PSF is degraded to the worst pawprint image at a given epoch.

The PSF magnitudes are placed on an absolute scale through a comparison with the VDFS magnitudes, which is calibrated as in \cite{carlos2018}, by applying systematic shifts of about 0.46 mag in $J$ and $0.25$ mag in $K_\mathrm{s}$ (see Appendix \ref{photshifts}), and are not corrected for reddening. We note that with this process the PSF on individual half-tile images, that refer to a set of observations obtained at a specific epoch during $\sim$1 hour, will be more homogeneous than that resulting from deep and stacked observations combining epochs observed during different sky conditions. While this represents a good average that can serve multiple-scientific applications, a dedicated processing tailored to address specific aspects, such as the determination of the SFH (Sec.\,\ref{sfh}) or the detection of semi-resolved star clusters \citep{miller2026}, may provide better results.

\section{Differences with respect to VMC}	
\label{difference}
Compared to previous VISTA observations of the Clouds obtained in the VMC survey \citep{cioni2011}, this programme adopts a larger number of jitter positions (9 versus 5 in VMC) and a reduced number of DITs (2 versus 8 and 10 versus 15 in VMC for the $J$ and $K_\mathrm{s}$ band, respectively). This results in less images per pawprint at a given position in the $J$ band (18  vs 40 in VMC, where the number of images is given by the product of the number of DITs and Jitters) and consequently a shorter exposure time per epoch (360 s versus 800 s in VMC). The opposite occurs in $K_\mathrm{s}$ (90 vs 75 images in VMC and  900 s versus 750 s exposure time in VMC). This programme uses only two filters ($J$ and $K_\mathrm{s}$) instead of three (VMC also observed in the $Y$ band). The exposure time was designed to detect faint pulsating variable stars in single-epoch images and old main-sequence turn-off stars in stacked images whereas the choice of filters reflects the best calibrated filters (VISTA photometry is calibrated with respect to 2MASS which does not have a $Y$ filter) and those where variable stars obey a clear, and narrow, period-magnitude relation. The centres of the tiles produced with this programme differ from those of the VMC survey. The deep VISTA tile in the LMC from this programme overlaps mostly with VMC tiles LMC 6\_4 and LMC 6\_5 whereas in the SMC the overlap is with tiles SMC 4\_3 and SMC 4\_4. 

The average parameters indicating the quality of the data, covering the same regions, are given in Table \ref{table:quality}. Compared to the VMC data, the FWHM of individual sources is about 0.2 arcsec smaller in both bands, which is larger than the standard deviations of the measurements. The difference in ellipticity is comparable to the uncertainties, but the dispersion is larger in this program than in the VMC survey, which could be due to an increased sensitivity to the source profile. The limiting magnitude, computed from the average of the limiting magnitude of individual pawprints, is similar in $K_\mathrm{s}$ and about 0.3 mag shallower in $J$ than in the VMC survey, due to the difference in exposure time between the two programmes, but the dispersion is lower. Note that when pawprints are mosaicked into tiles the sensitivity increases by $\sim$0.3 mag due the detectors overlaps and further due to the stacking of multiple epochs. The significantly larger number of epochs (about 40 in both $J$ and $K_\mathrm{s}$) obtained with this programme compared to those in the VMC survey (3 in $J$ and 12 in $K_\mathrm{s}$), combined with high-quality sky conditions, produces the best near-infrared survey for the centres of the Clouds to-date.

	\begin{table}
	\setlength{\tabcolsep}{3pt}
		\caption{Quality of the data compared to VMC data.}                        
		\label{table:quality}      
		\small
		\begin{tabular}{lcccc}
			\hline \hline
			Survey & Band & FWHM & Ellipticity & Sensitivity \\
			 & & ($^{\prime\prime}$) & & (mag) \\
            \hline
            VMC & J & 0.99$\pm$0.11 & 0.06$\pm$0.01 & 19.80$\pm$0.40 \\
             & K$_\mathrm{s}$ & 0.93$\pm$0.10 & 0.05$\pm$0.01 & 18.72$\pm$0.28 \\
            This work & J & 0.79$\pm$0.08 & 0.08$\pm$0.04 & 19.53$\pm$0.16 \\ 
             & K$_\mathrm{s}$ & 0.75$\pm$0.07 & 0.07$\pm$0.05 & 18.68$\pm$0.10 \\
			\hline
		\end{tabular}
		\tablefoot{The limiting magnitude is the average from individual pawprints.}
	\end{table}

\section{Content of the catalogue}
\label{quality}

The number of unique detections resulting from the deepstack images in $J$ and $K_\mathrm{s}$ are given in Table \ref{sources}. There are in total about 2.5 million detections equally split between the SMC and the LMC. The majority (80\%) are sources detected in two filters of which 2/3 have a star-like profile and 1/3 a galaxy-like profile. An approximately equal number of sources are detected only in one filter and 0.3\% are due to noise or saturated objects. In the SMC, there are more stars detected only in $J$ than in $K_\mathrm{s}$, while the opposite is true in the LMC. The number of extended sources in the SMC is larger than that of stars whereas in the LMC they are not significantly different. Note also that sources in detector \#16 appear as single $K_\mathrm{s}$-band detections because the variability of the quantum efficiency is worse in the $J$ band, which impacts the flatfielding process and the ability to detect sources on the affected images. In what follows, we only refer to sources with minor quality issues ({\it ppErrBits}<256). This criterium excludes sources that lie within detection \#16 or an underexposed area of a tile as well as within a dithered offset of a stacked frame boundary. It also excludes sources close to saturation, with a problematic photometric calibration or that while appearing in a tile they are not detected in the corresponding pawprints. At least 80\% of the sources detected in two filters satisfy this criterium. In the SMC, this is also true for single detections in $K_\mathrm{s}$ whereas in $J$ they drop to 65\%. In the LMC, there are more good-quality sources detected only in $J$ (75\%) than in $K_\mathrm{s}$ (50\%).

\begin{table}
	\setlength{\tabcolsep}{3pt}
	\caption{Number of detections.}
	\label{sources}
	\centering
	\small
	\begin{tabular}{llrrrrr}
	\hline\hline
	 Galaxy & Filter(s) & Star & Galaxies & Stars$^\mathrm{a}$ & Galaxies$^\mathrm{a}$ &Others$^\mathrm{b}$ \\
	\hline
	 SMC & $J+K_\mathrm{s}$ & 648 000 & 257100 & 543 000 & 199 800 & 400 \\
	 & $J$ only & 63 800 & 101 100 & 42 600 & 63 100 & 2 100 \\
	 & $K_\mathrm{s}$ only & 24 300 & 68 700  & 19 800 & 50 500 & 1 000 \\
	 & All & 736 100 & 426 900 & 605 400 & 313 400& 3 500 \\
	 & & & & & & \\
	 LMC & $J+K_\mathrm{s}$ & 789 800 & 273 500 & 648 800 & 215 100 & 300 \\
	 & $J$ only & 38 000 & 56 200 & 29 800 & 39 800 & 1 500 \\
	 & $K_\mathrm{s}$ only & 59 000 & 58 300 & 25 600 & 35 000 & 1 500 \\
	 & All & 886 800 & 388 000 & 704 200 & 289 900 & 3 300 \\
	 \hline
	\end{tabular}
	
	\tablefoot{Subscript $^\mathrm{a}$ refers to the subset of sources with minor quality issues ({\it ppErrBits}<256) and subscript $^\mathrm{b}$ refers to detections with a 90\% probability of being noise or saturated sources.}
	
\end{table}

\begin{figure}
\centering
\includegraphics[width=\hsize]{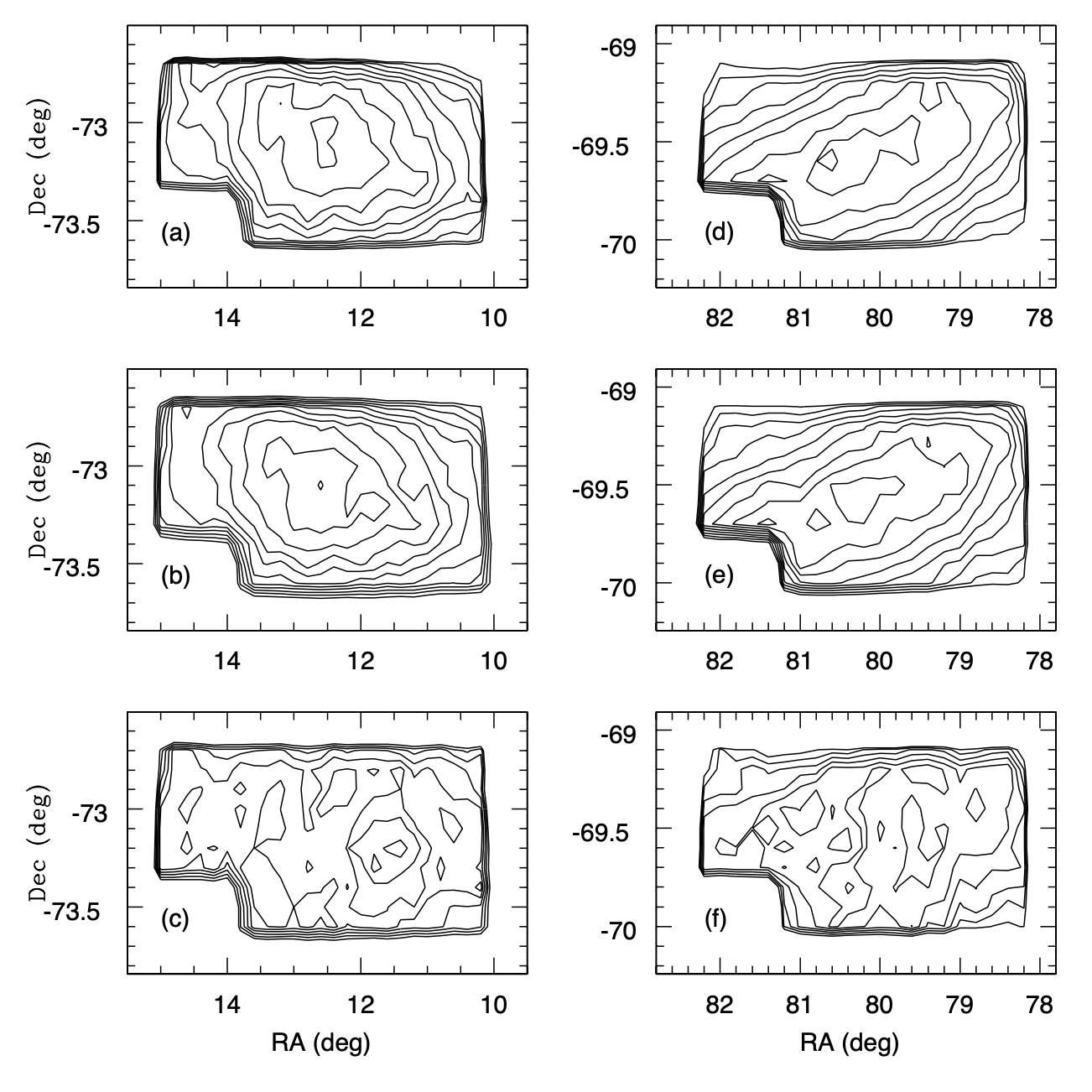}
\caption{Spatial distribution of sources from the SMC (left) and LMC (right) catalogue detected in both $J$ and $K_\mathrm{s}$, and with {\it ppErrBits}<256. Top panels refer to all sources, middle panels refer to stars and bottom panels refer to galaxies. Contours mark number density levels increasing towards the centres as follows. Panel (a): from 2 000 in steps of 200. Panel (b): from 1 000 in steps of 200. Panel (c): from 500 in steps of 75. Panel (d): from 3 000 in steps of 300. Panel (e): from 2 000 in steps of 250. Panel (f): from 800 in steps of 75. The missing bottom-right corner corresponds to the area covered by detection \#16.}
\label{maps}
\end{figure}

The spatial distributions of sources within the central region of each galaxy are shown  in Figure \ref{maps}. They contiguously cover the central regions of the galaxies, contrary to the VMC survey (Fig.~\ref{vmcmaps}), which clearly favours the characterisation of morphological structures. 
They illustrate the expected increase in the number of sources towards the centres. The distribution of galaxies is unstructured, also as expected, whereas stars depict the elongations of the SMC from the south-east to north-west and along the LMC bar. 

\begin{figure}
\centering
\includegraphics[width=0.9\hsize]{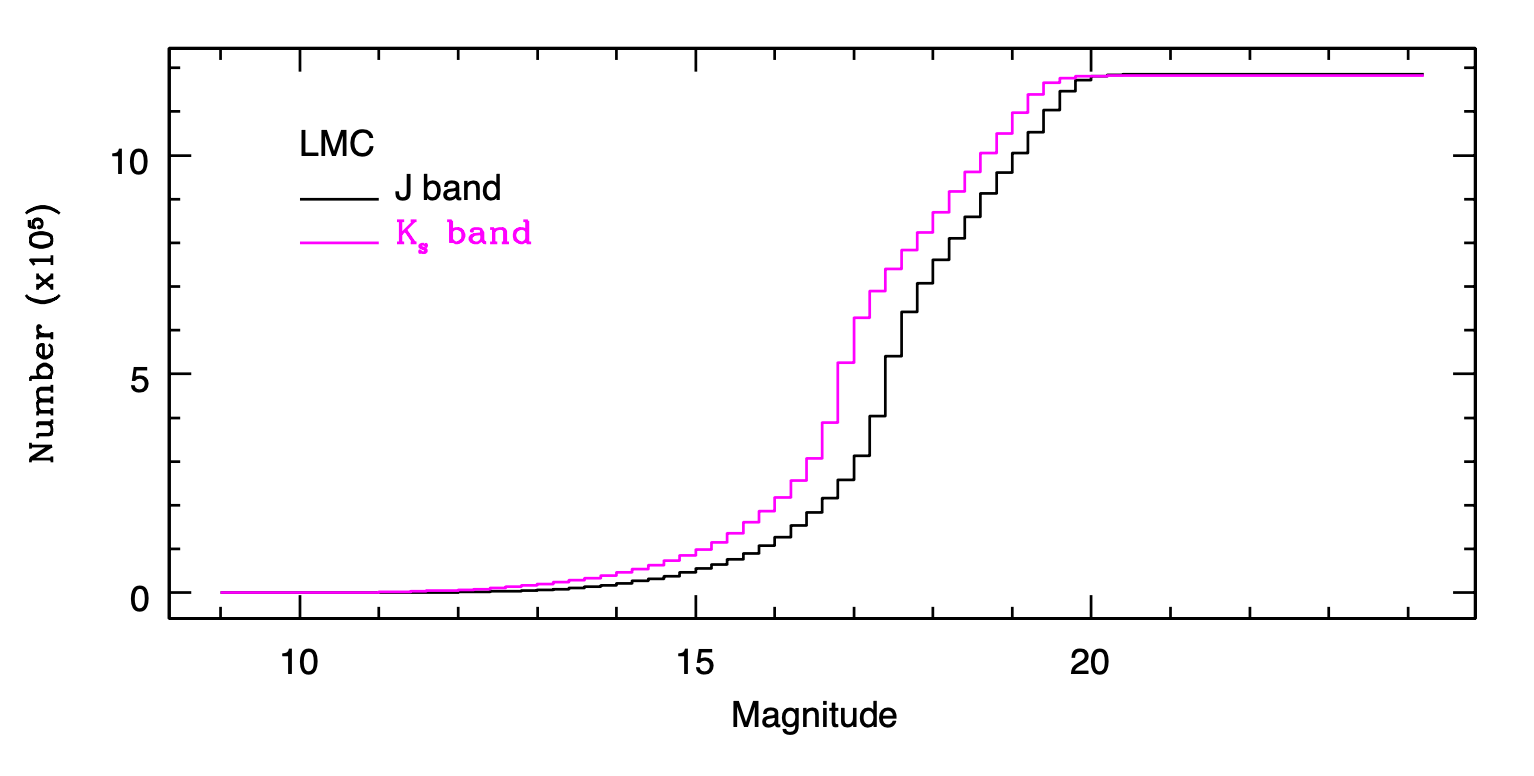}
\includegraphics[width=0.9\hsize]{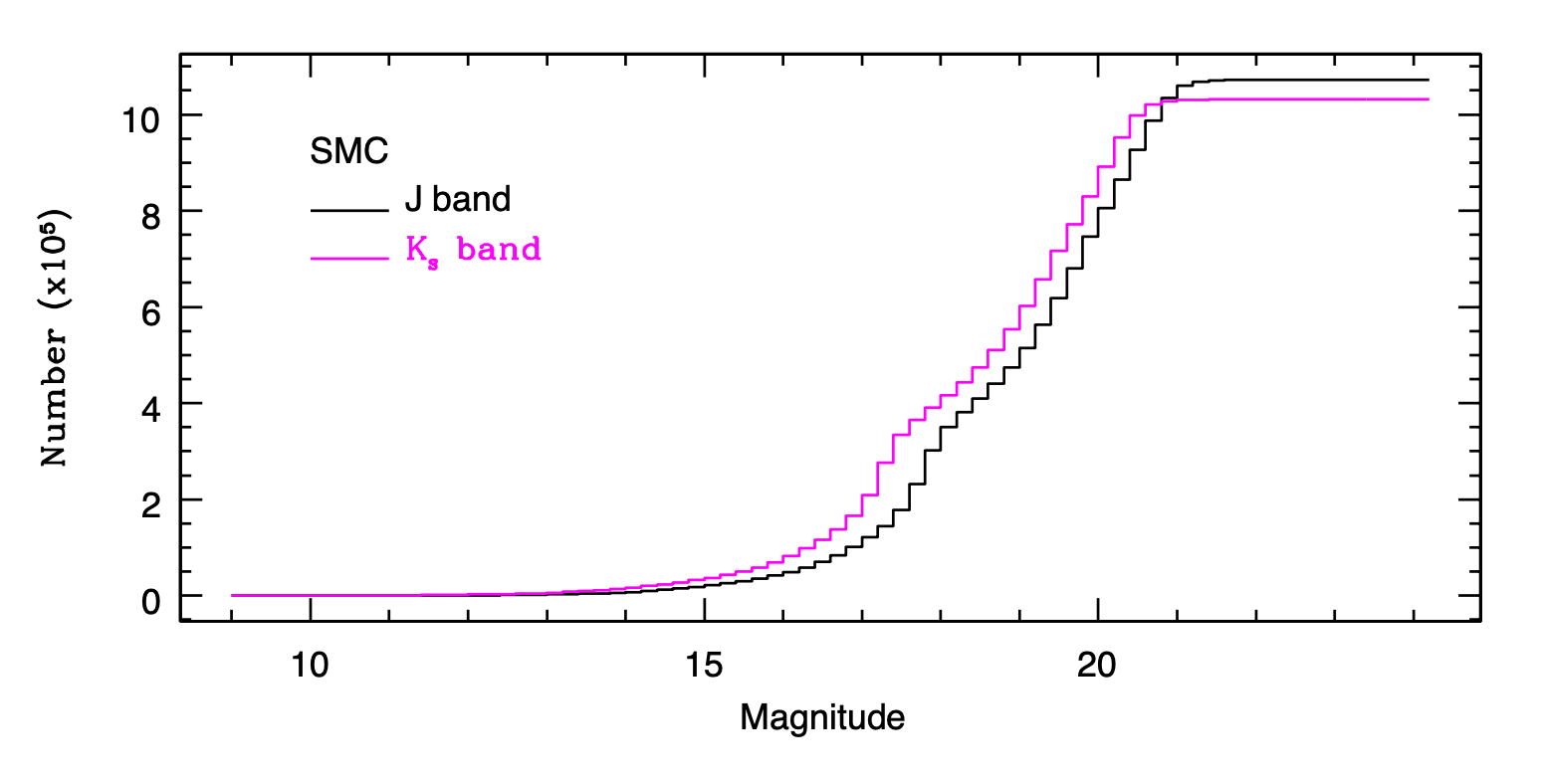}
\caption{Cumulative star counts in bins of 0.2 mag for the entire LMC (top) and SMC (bottom) catalogues in the $J$ (black) and $K_\mathrm{s}$ (magenta) bands.}
\label{cumulative}
\end{figure}

\begin{figure*}
\centering
\includegraphics[width=9cm]{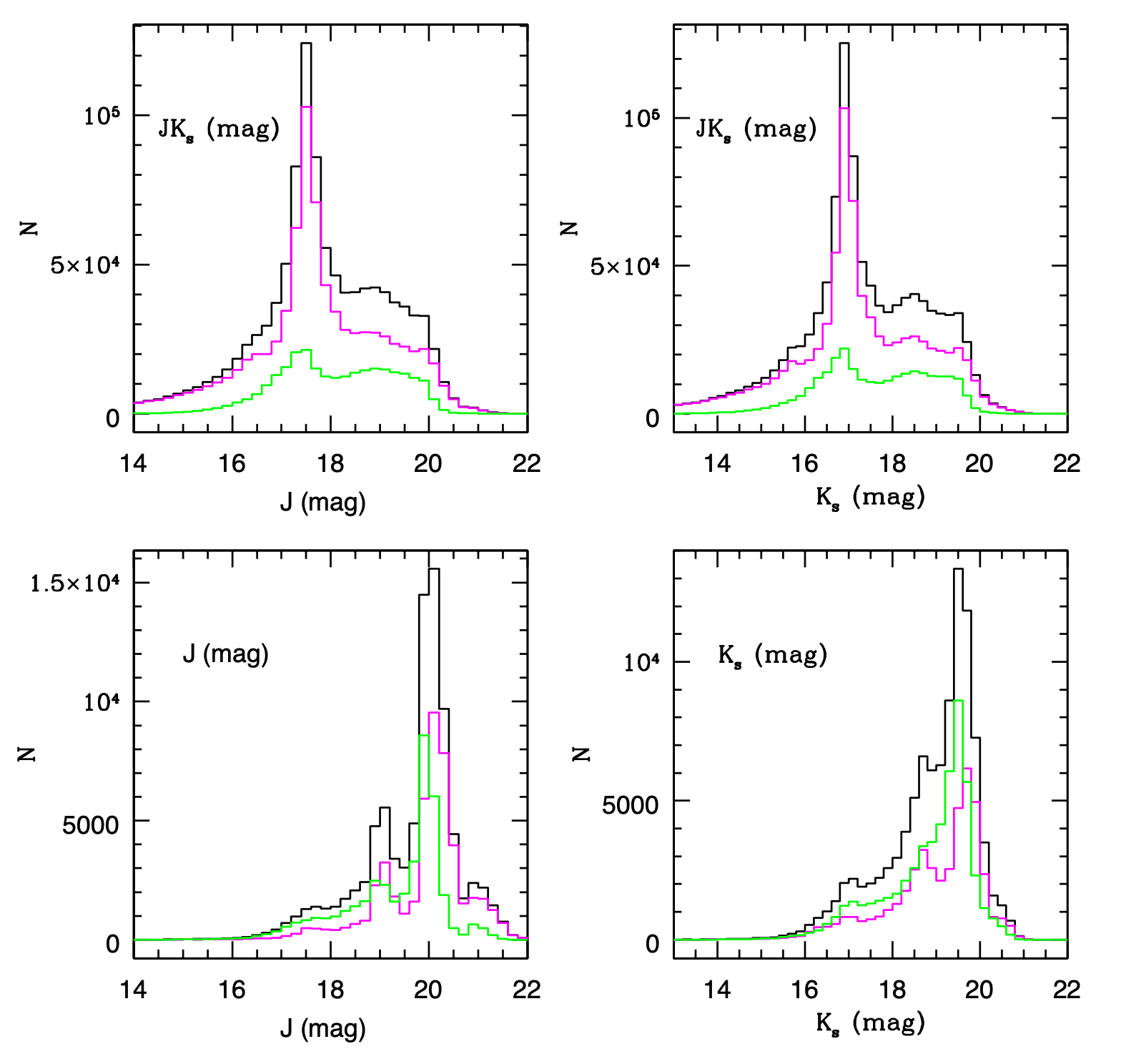}
\includegraphics[width=9cm]{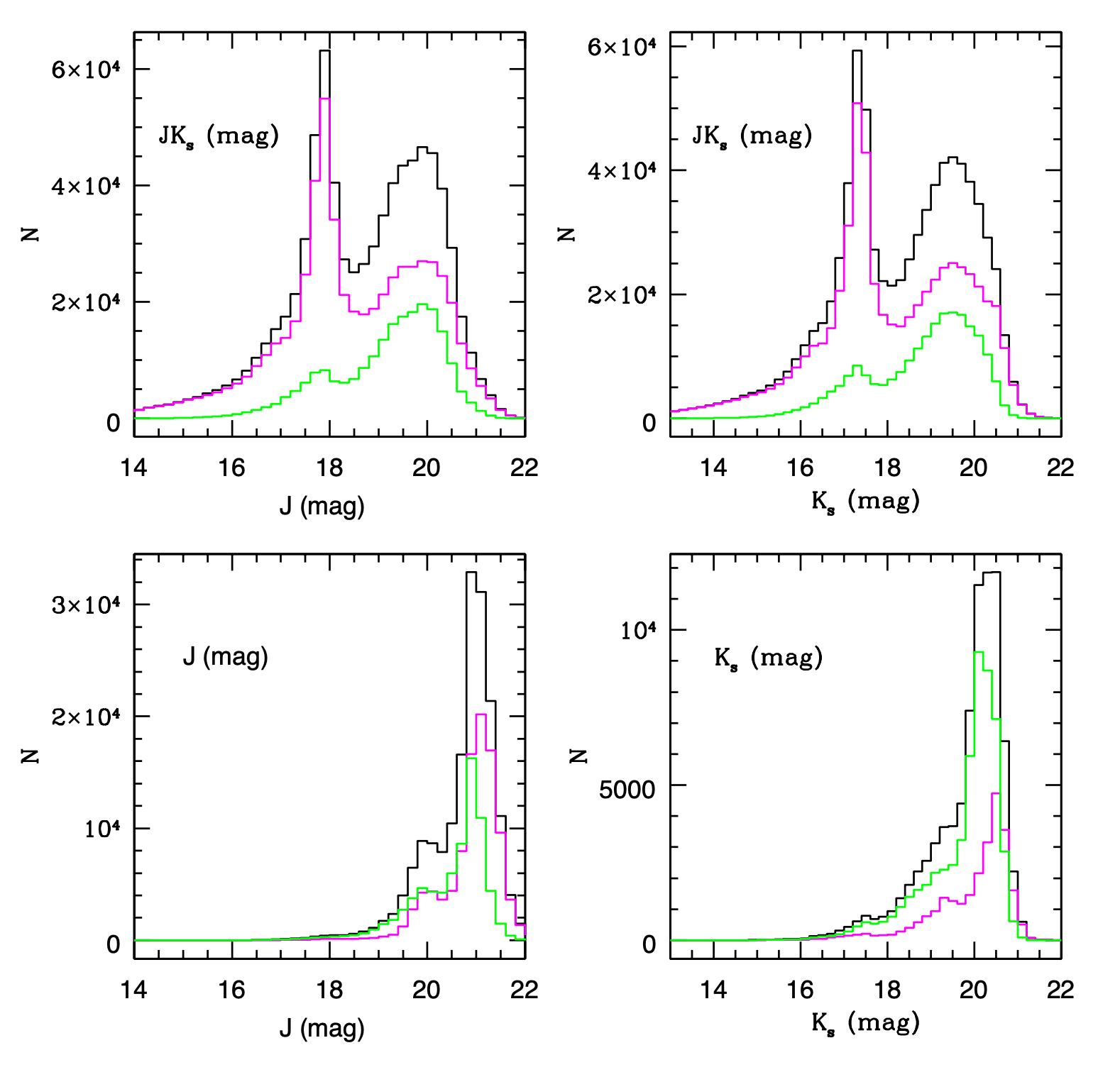}
\caption{Apparent luminosity function in 0.2 mag bins, for sources with {\it ppErrBits}<256, with detections in one (bottom) and two (top) filters, as indicated on the upper left corner of each panel. Sources with a stellar profile are shown in magenta and sources with a galaxy profile are shown in green whereas their total is shown in black. The four panels on the left refer to the LMC and those on the right to the SMC.}
\label{histograms}
\end{figure*}

The number of sources increases as a function of magnitude as shown in Figure \ref{cumulative}. It reaches a plateau at $J\sim20.5$ mag and $K_\mathrm{s}\sim20.3$ in the SMC and about half a magnitude brighter in the LMC, which hint at the approximate completeness of the catalogue. The absolute completeness will be quantified using artificial stars tests in the study of the SFH (Sect.\,\ref{sfh}). Figure \ref{histograms} shows the apparent luminosity functions. The highest peaks in the distributions of stellar sources detected in both filters are created by red clump stars while the second highest peaks are due to RGB stars. Detections in the SMC reach sources $\sim$1 mag fainter than in the LMC. Many of the sources with a galaxy-like profile (galaxies hereafter) are likely stellar blends because compared to less dense stellar regions, where background galaxies are more easily distinguished from stars, the background galaxies dominate for $J$>20 mag and $K_\mathrm{s}$>19 mag (cfg. with Fig.\,2 in \citealp{cioni2025}). The number of background galaxies in the high stellar density regions along the LMC bar can drop to 10--20 galaxies per detector \citep{niederhofer2022}.

\begin{figure*}
\centering 
\includegraphics[width=8.9cm]{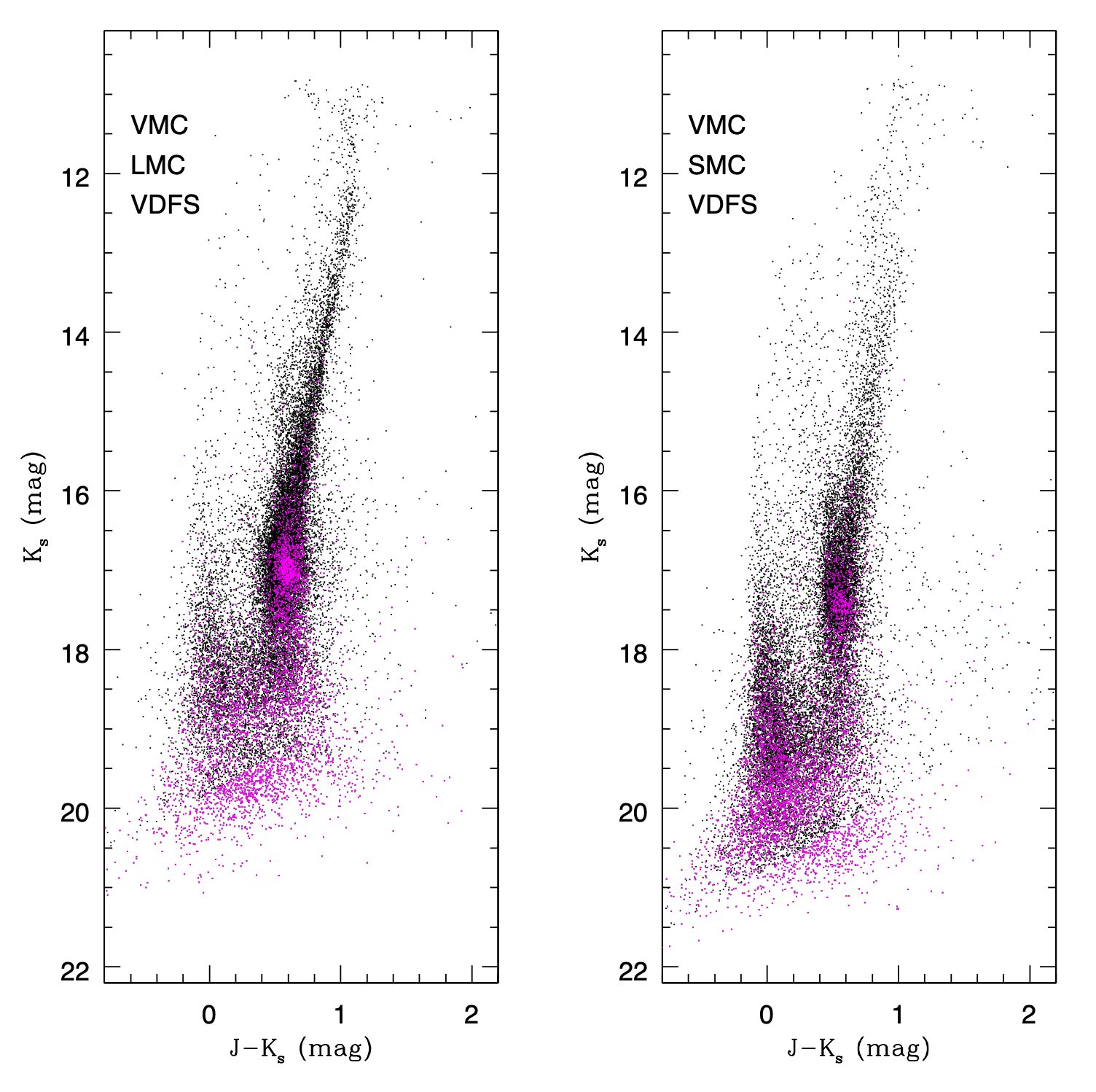}
\includegraphics[width=9cm]{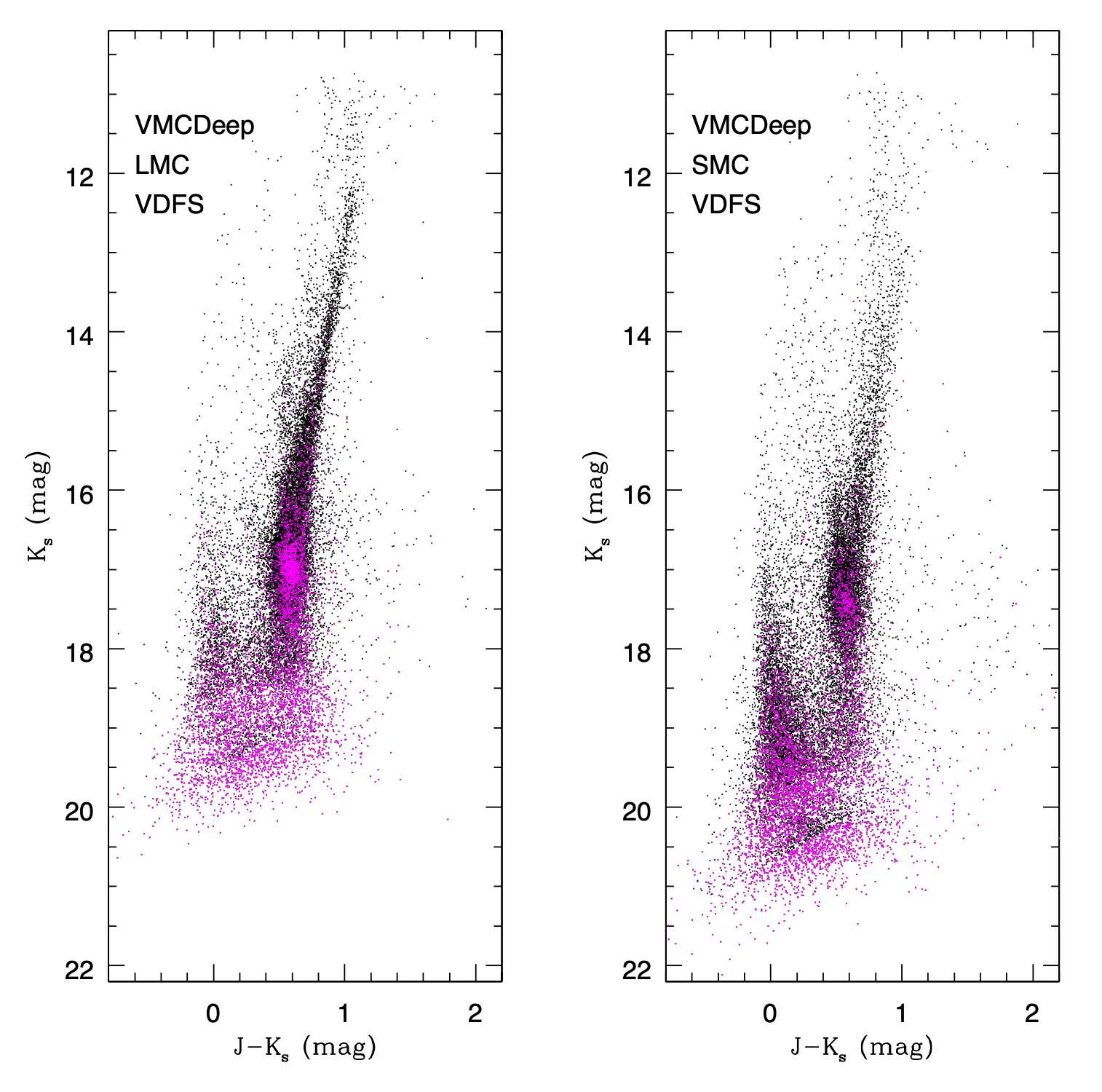}
\includegraphics[width=9cm]{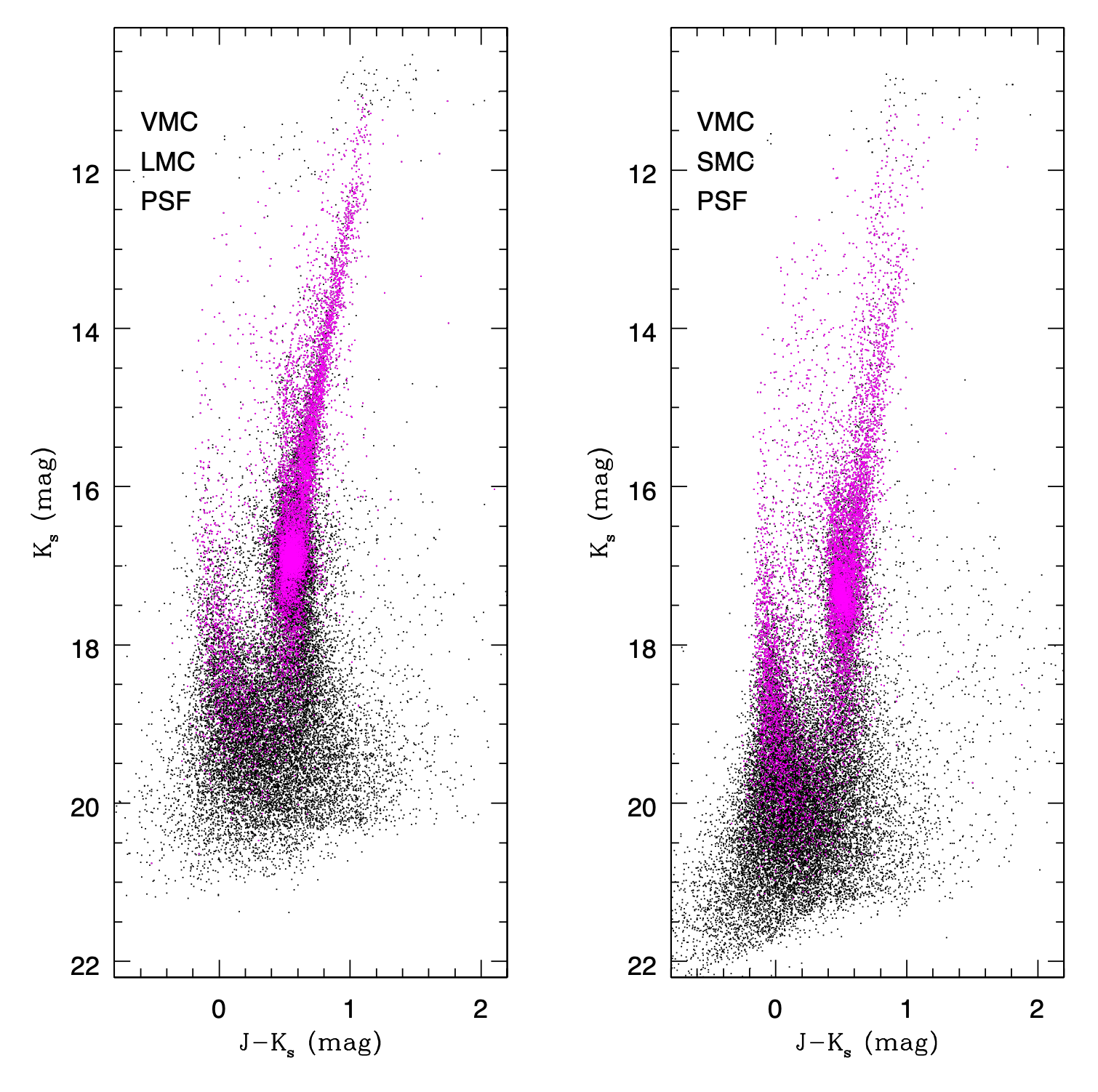}
\includegraphics[width=9cm]{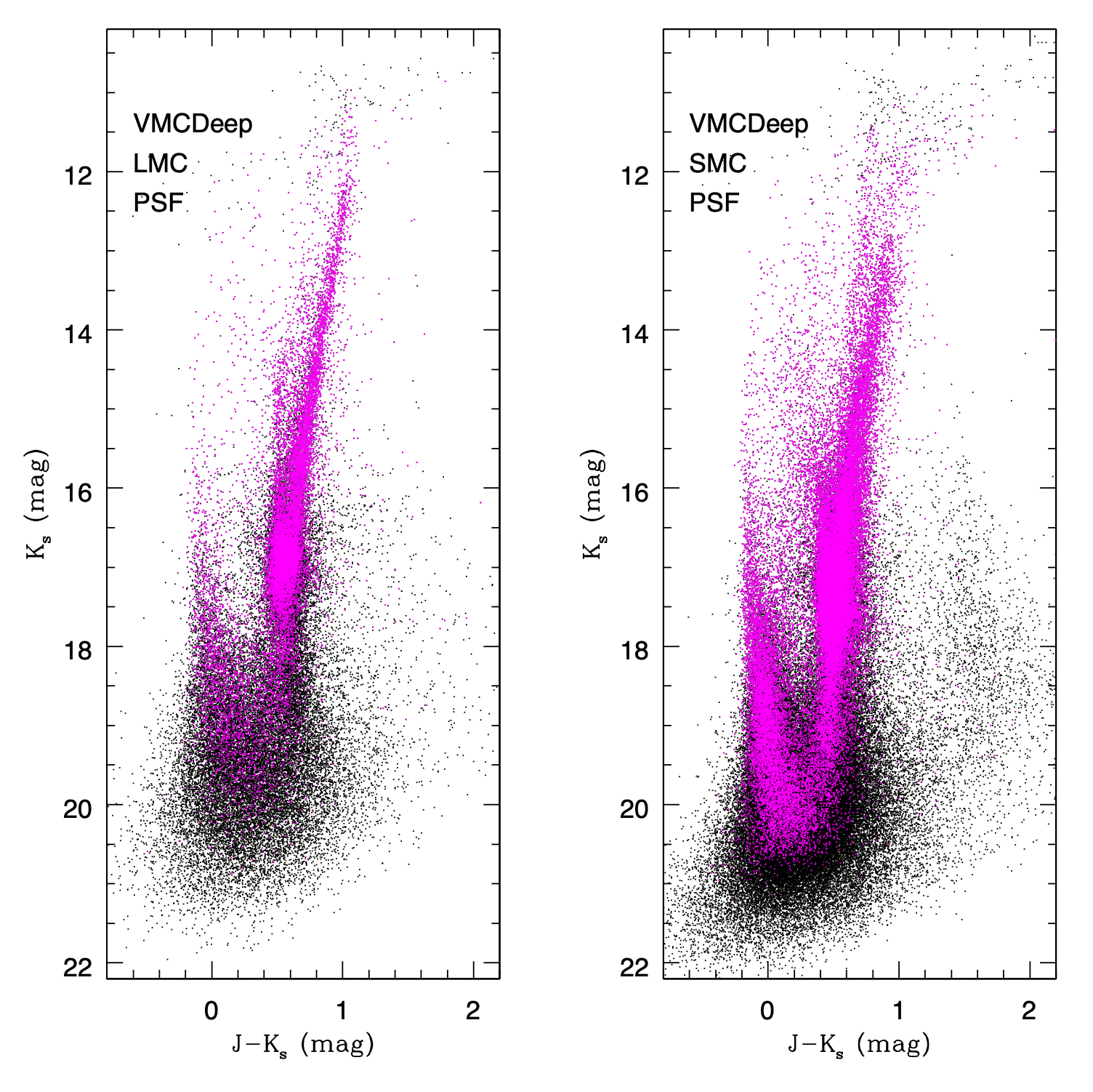}
\caption{Colour-magnitude diagram of sources in the central 0.1 deg$^2$ of the LMC and SMC comparing detections in the VMCDeep and VMC survey programmes. The top panels show all sources detected by the VDFS (black) and only those with {\it ppErrBits}=0 (magenta), corresponding to high-quality isolated sources. The bottom panels show all sources detected with PSF photometry (black) and only those that show SHARP values between --0.3 and +0.3 (magenta), indicating well-defined stellar objects.}
\label{cmd}
\end{figure*}

Figure \ref{cmd} shows the distribution of sources located the central 0.1 deg$^2$ region of the LMC and SMC in the colour-magnitude diagram. In this figure, sources from the VMCDeep and VMC catalogues are compared as well as sources extracted using the VDFS and PSF photometry. As expected, there are more sources with PSF than VDFS photometry. There are also more sources in the LMC than in the SMC, but in the SMC there are more faint sources, probably due to the lower stellar density than in the LMC. Considering only VDFS-based detections, which refer to magnitudes computed using apertures of 2 arcsec in diameter {\it (MagAper3}), there are $\sim$2,000 sources less in VMCDeep than in VMC. However, there are $\sim$1,000 more sources in VMCDeep than in VMC which have {\it ppErrBits}=0. These are sources of excellent quality from a detection point of view because they do not present any problem, within the aperture, and are sufficiently isolated from neighbouring sources. This means that sources in VMCDeep are better defined than in VMC. A similar conclusion is derived from the PSF photometry where there are 2--4 times more sources in VMCDeep than in VMC, depending on galaxy, and at least 20\% of them have SHARP values between $-$0.3 and +0.3. Note that to compare the PSF photometry we corrected for  systematic shifts between the VMCDeep and VMC catalogues which are slightly different from those applied to calibrate the PSF catalogue with respect to the corresponding VDFS one (see Sect.\,\ref{photshifts} for details).

\begin{figure*}
\centering
\includegraphics[width=8cm]{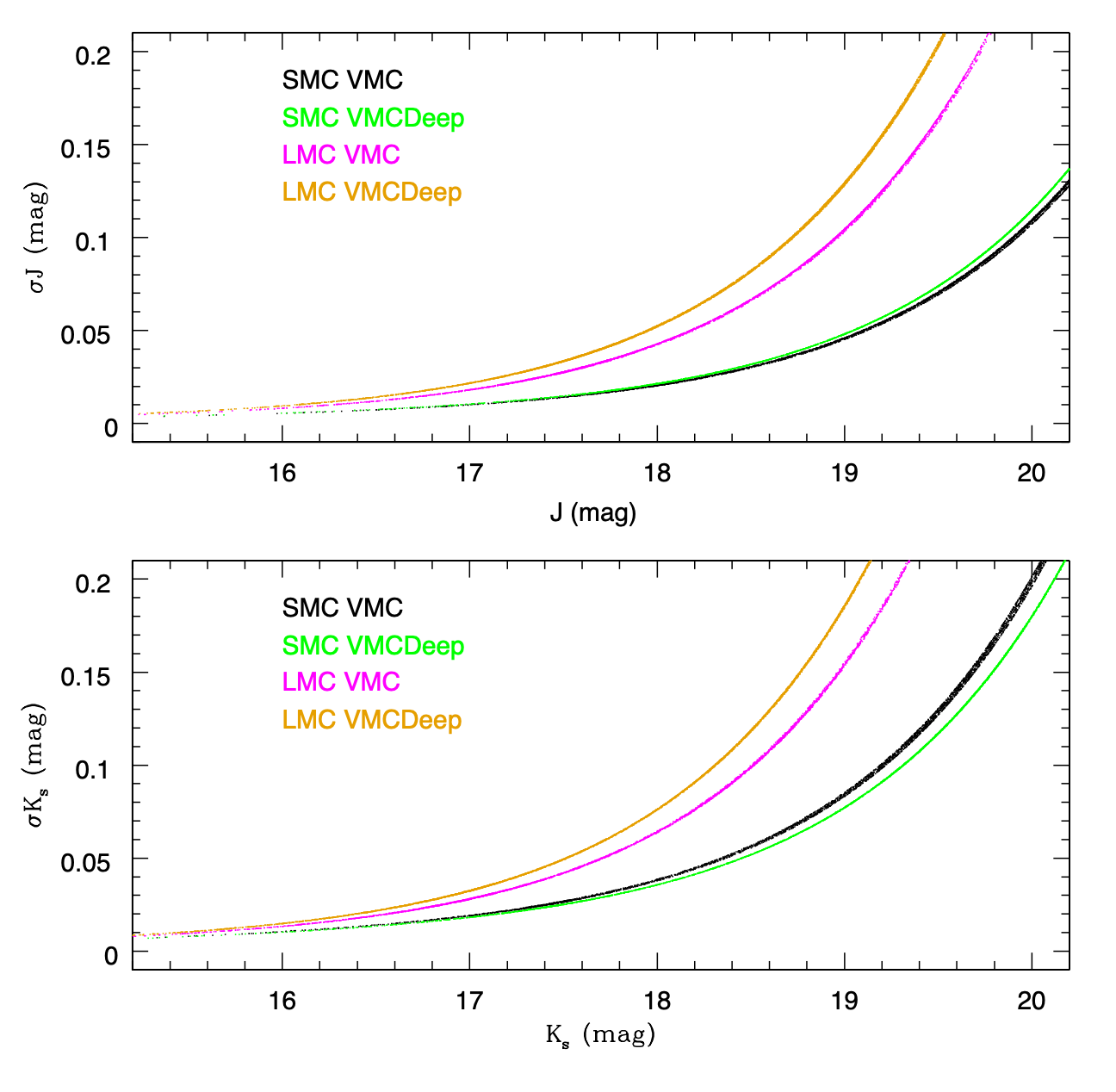}
\includegraphics[width=8cm]{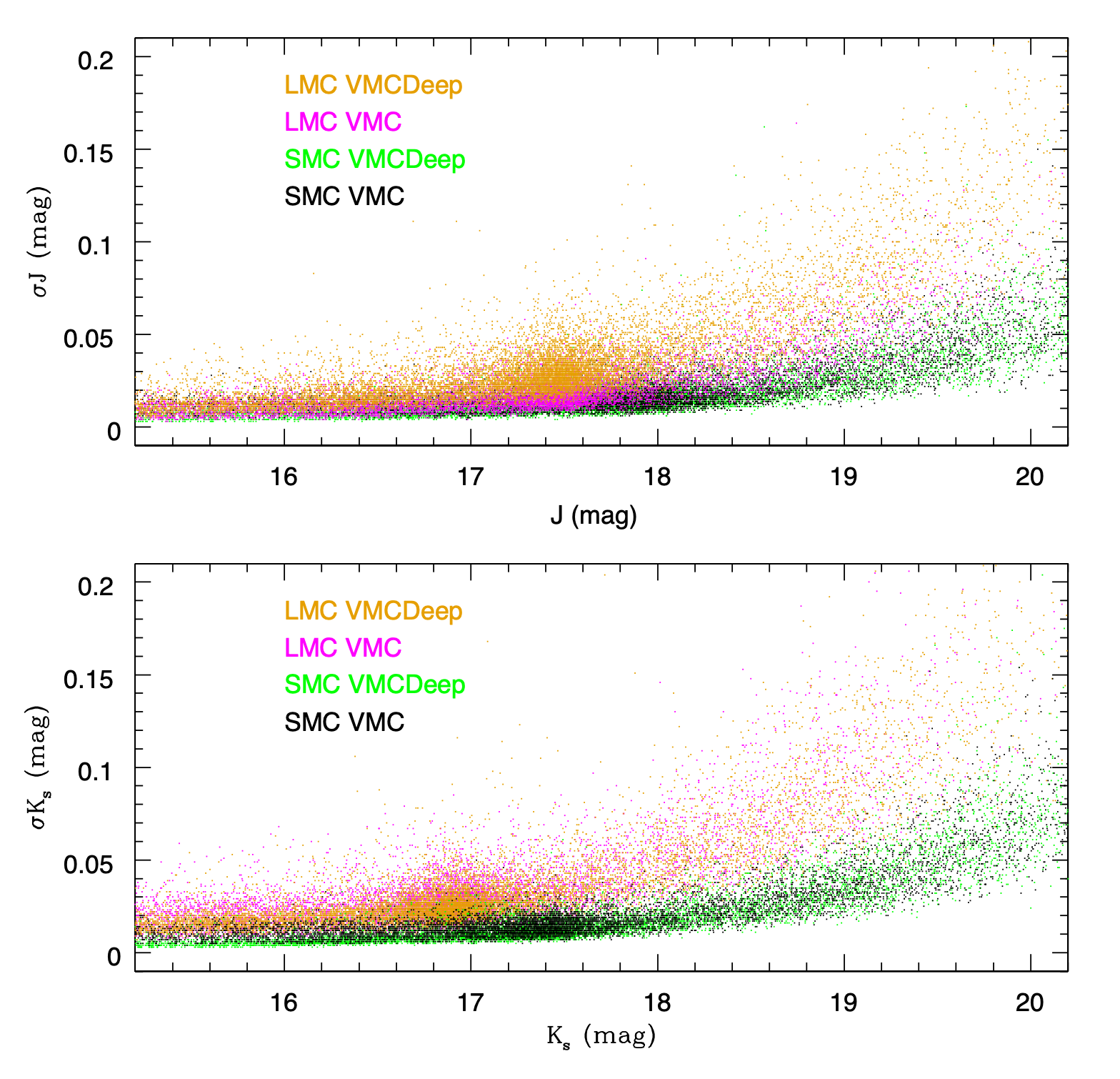}
\caption{Distribution of photometric uncertainties between this programme and the VMC survey as a function of magnitude. (Right) Sources detected in both bands using VDFS. (Left) Sources detected using PSF photometry where for clarity we only show those detected in both bands with SHARP values between --0.3 and +0.3. Colours refer to different samples as indicated within each panel.}
\label{differences}
\end{figure*}

The photometric uncertainties from this programme compared to the VMC survey, as derived from the VDFS, are shown in (Fig.~\ref{differences}). In the SMC, they are very similar in both bands, with smaller values for the $K_\mathrm{s}$ detections in VMCDeep than in VMC. In the LMC, the uncertainties are systematically larger in VMCDeep than in VMC. This decrease in photometric precision is likely driven by two compounding factors: the shallower depth of the individual pawprints (Sect.\,\ref{difference}) and the use of the same photometric aperture as in VMC, which for intrinsically smaller detections (Tab.\,\ref{table:quality}) degrades the S/N by incorporating excess background emission and blended flux from adjacent stars. By comparing the uncertainties obtained in single pawprints, using apertures of 1.4 arcsec ({\it aperMag2}) and 2 arcsec ({\it aperMag3}) from the vmcdeepDetection tables, we find a difference of $\sim$0.01 mag, which corroborates the previous statement. No systematic trends with colour are present. Using PSF photometry, we obtain similar uncertainties between both programmes also for the LMC and the remaining differences in the $J$ band are negligible. These will be further characterised in the study of the SFH (Sect.\,\ref{sfh}).

\subsection{Confusion}
\label{sec:confusion}
The improved spatial resolution of the VMCDeep observations combined with the increase in integration time, with respect to the VMC survey, has improved the clarity of the detections. To show this, we computed a qualitative confusion limit by calculating the number of beams per source in regions 0.2$\times$0.1 arcsec$^2$, where one beam corresponds to the area in arcsec$^2$ occupied by a source with radius equal to half the FWHM. Using the FWHM values in Tab.\,\ref{table:quality} we obtain the maps shown in Fig.\,\ref{confusiondeep}; similar maps obtained for the VMC survey catalogue \citep{cioni2025} are shown in the Appendix (see also \citealp{vijayasree2026}). A critical limit corresponds to 20--25 beams, up to 50 beams confusion is high, whereas above 50 beams is low. Confusion can influence both the magnitude and position of faint sources as well as their number. A low confusion level means a high completeness and reliable source detection. 

\begin{figure}
    \centering
    \includegraphics[width=\hsize]{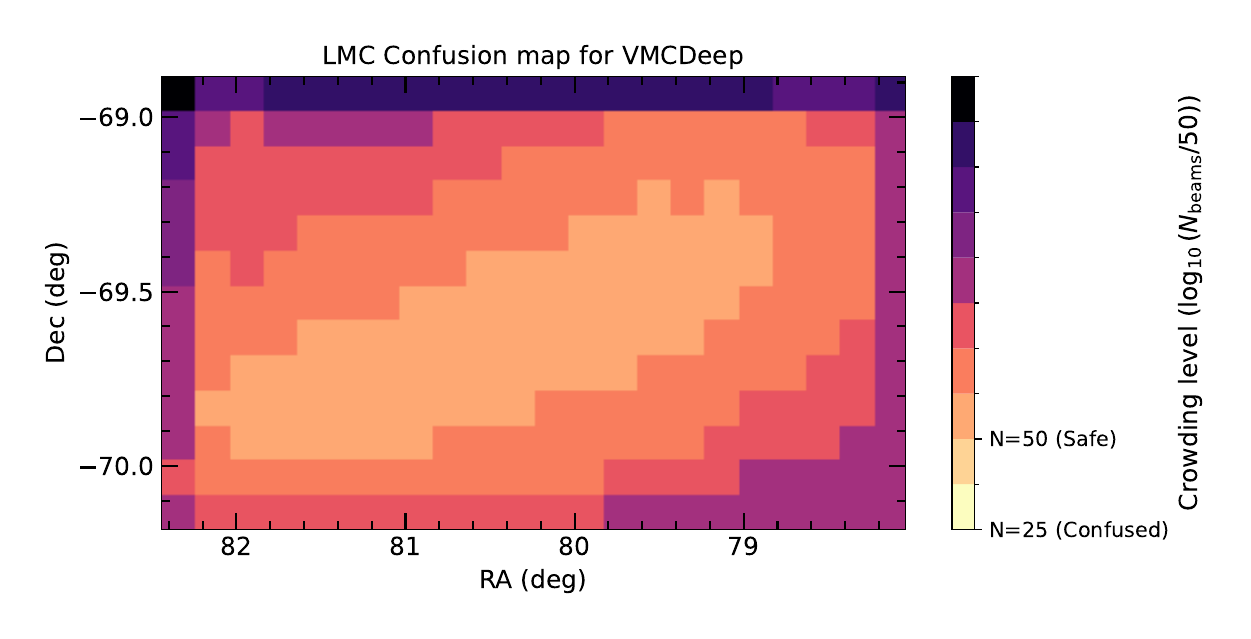}
    \includegraphics[width=\hsize]{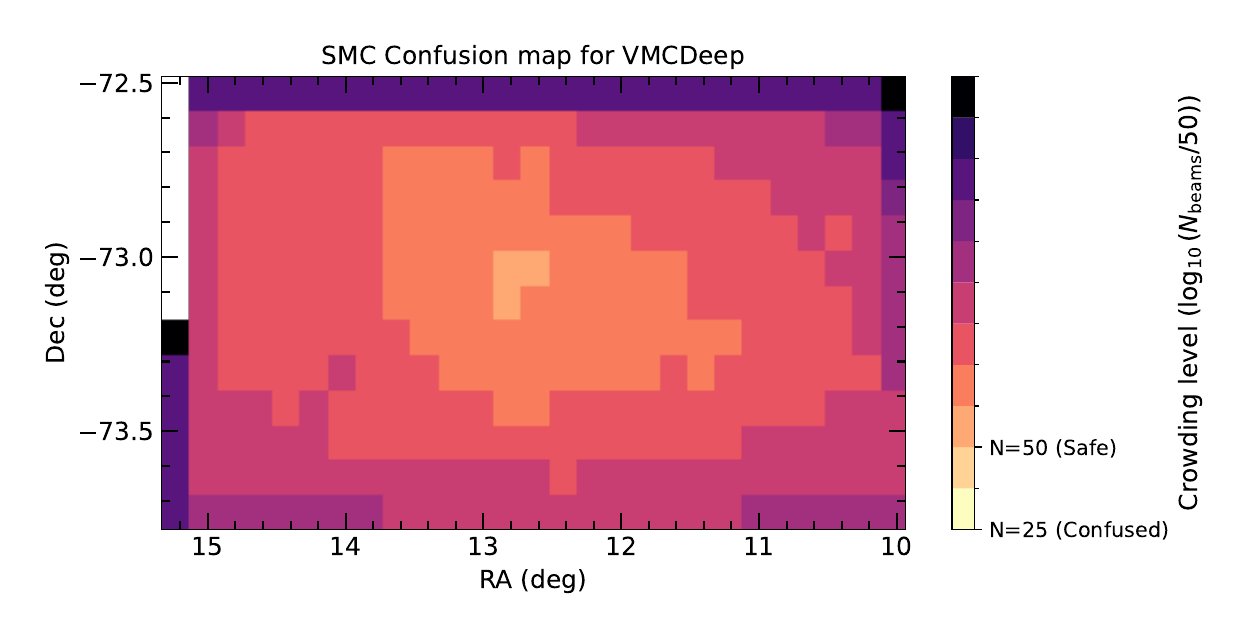}
    \caption{Confusion maps for bins of 0.2$\times$0.1 deg$^2$ in the LMC (top) and SMC (bottom) central regions. The threshold number of beams (N=50) and the critical value (N=25) are indicated in the colour bar.}
        \label{confusiondeep}
\end{figure}

\subsection{Data availability}
\label{public}
Data products resulting from both the CASU and WFAU reductions, which are based on the VDFS, are stored in the VSA. They are made publicly available to the astronomical community through the VSA and the ESO Science Archive Facility (SAF; \citealp{romaniello2023}). Catalogues with PSF photometry are not released at this stage because they are still undergoing validation by the science team, which occurs through the application of the catalogues to specific science analysis. They will be made available together with the first respective publications.

\section{Applications of the catalogue}
\label{content}

\subsection{Stellar kinematics}
\label{kinematics}

The main goal of this programme is to constrain the central stellar kinematic patterns of the Clouds through measurements of near-infrared proper motions with better quality observations and spanning a longer time baseline in combination with the VMC survey. In these regions, the VMC survey obtained on average 12 epochs in the $K_\mathrm{s}$ band over a time baseline of two years (\citealp{niederhofer2021,niederhofer2022}). One additional epoch was acquired several years later resulting in a time baseline of $\sim$7 years for the SMC and $\sim$11 years for the LMC central regions \citep{vijayasree2025}. The VMC survey obtained only a few observations in the $J$ band within the main programme and there were no additional ones that followed them. The VMCDeep programme provides a significant increase in the number of epochs (at least 37 or 40 in the SMC or LMC) and an increase in the time baseline of another 1--3 years, depending on which of the two VMC tiles covering the centres of each galaxy are considered (Sect.~\ref{observations}). 

\begin{figure}
    \centering
    \includegraphics[width=8cm]{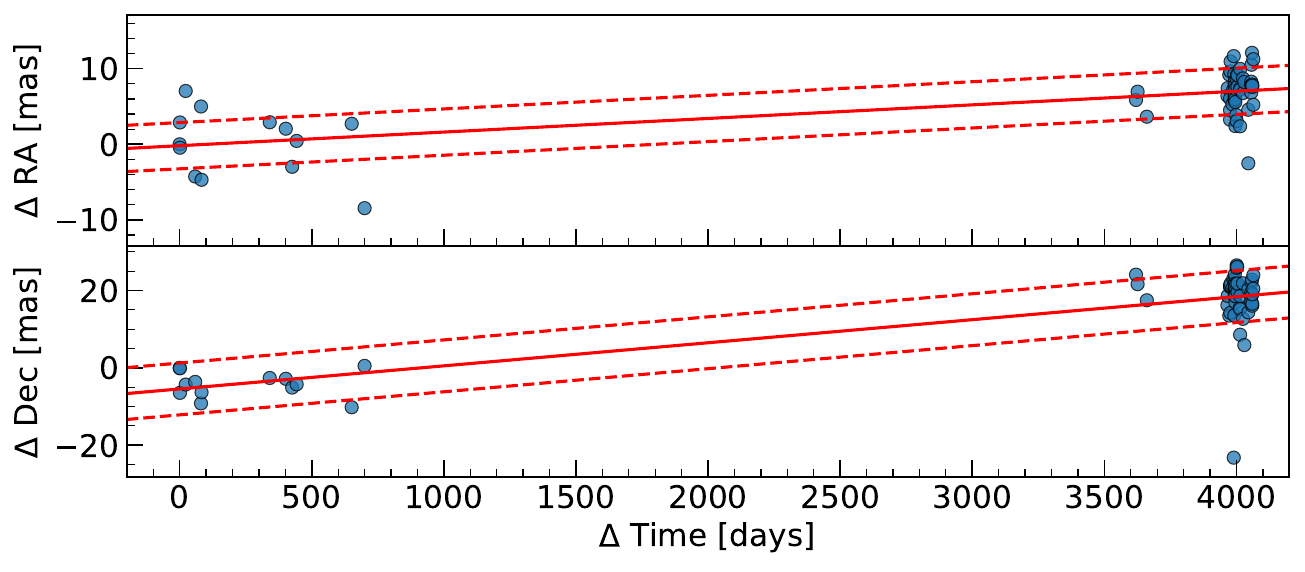}
    \includegraphics[width=8cm]{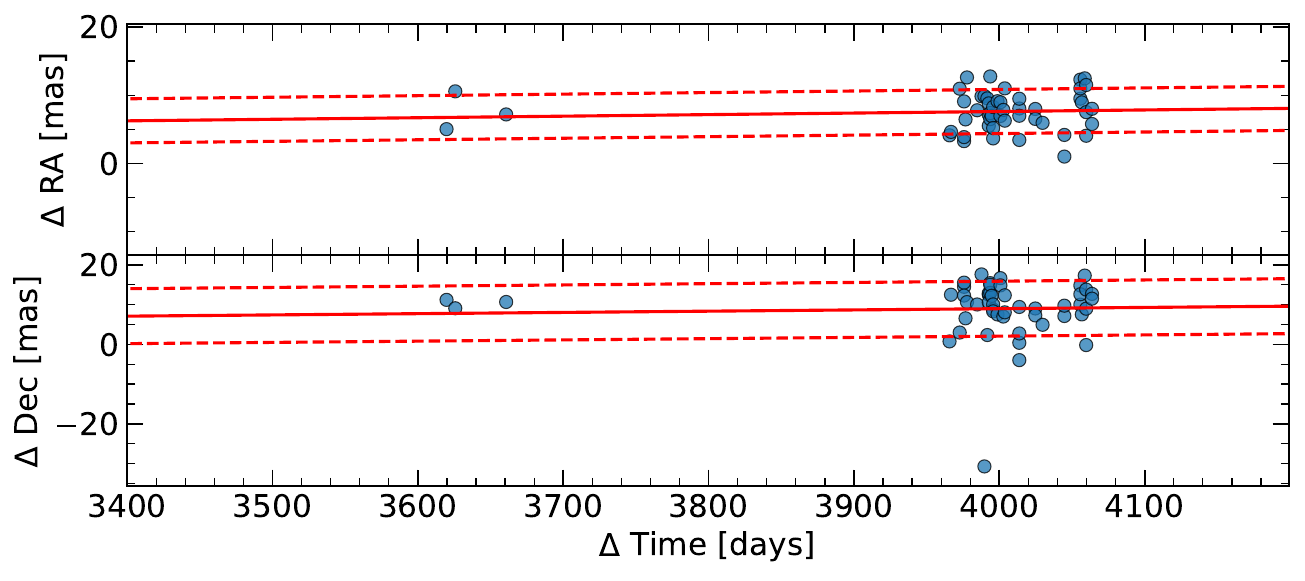}
    \caption{Sky position of a single random star as a function of time observed in the standard VMC survey ($\Delta$Time<1000 days), in additional epochs ($\Delta$Time$\sim$3600 days) and in this programme ($\Delta$Time$\sim$4000 days). The bottom panel is a zoom-in version of the top panel. Both good and bad quality epochs are shown for each programme.  The solid line is the best-fitting linear curve to the data. The dashed lines indicate the $\sigma$ distance of the stars from the fitted line.}
        \label{propermotion}
\end{figure}

Figure \ref{propermotion} shows the variation of the celestial coordinates for a random star observed in all three programmes. The slopes of the variations, which correspond to the proper motions in the two directions, are significantly better constrained in the combined data set than in the VMC or VMCDeep programme alone. Preliminary results based on proper motions derived from both bands show that the proper motion precision improves by about 2.5 times, compared to previous studies. In addition, using only sources detected in at least 70\% of the epochs adds fainter stars (by about half a magnitude; VMC-only proper motions are limited to sources with $K_\mathrm{s}<18.5$ mag, e.g.\,\citealp{niederhofer2022}) to the sample of usable stars for statistically addressing the kinematical pattern of the central regions, without significantly diminishing the proper motion precision.

\subsection{Variable stars}
\label{variables}

\begin{figure}
    \centering
    \includegraphics[width=8cm]{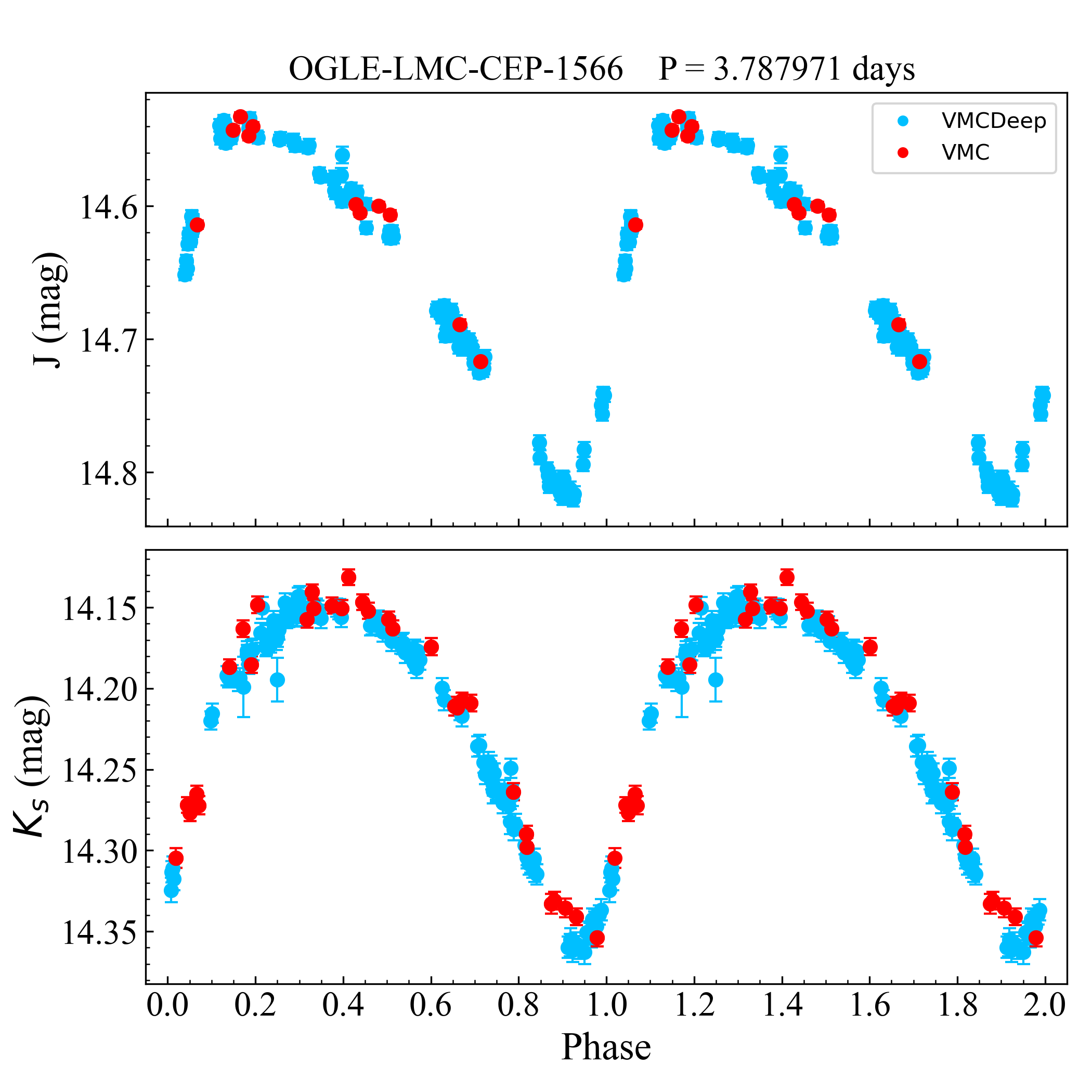} 
    \caption{VISTA light-curve of a Classical Cepheid in the $J$ (top) and $K_\mathrm{s}$ (bottom) bands. Data are folded according to the period and the epoch of maximum light derived from OGLE. The different colours indicate detections from this programme (light blue) and from VMC DR7 (red).}
         \label{lightcurve}
\end{figure}

\begin{figure}
    \centering
    \includegraphics[width=8cm]{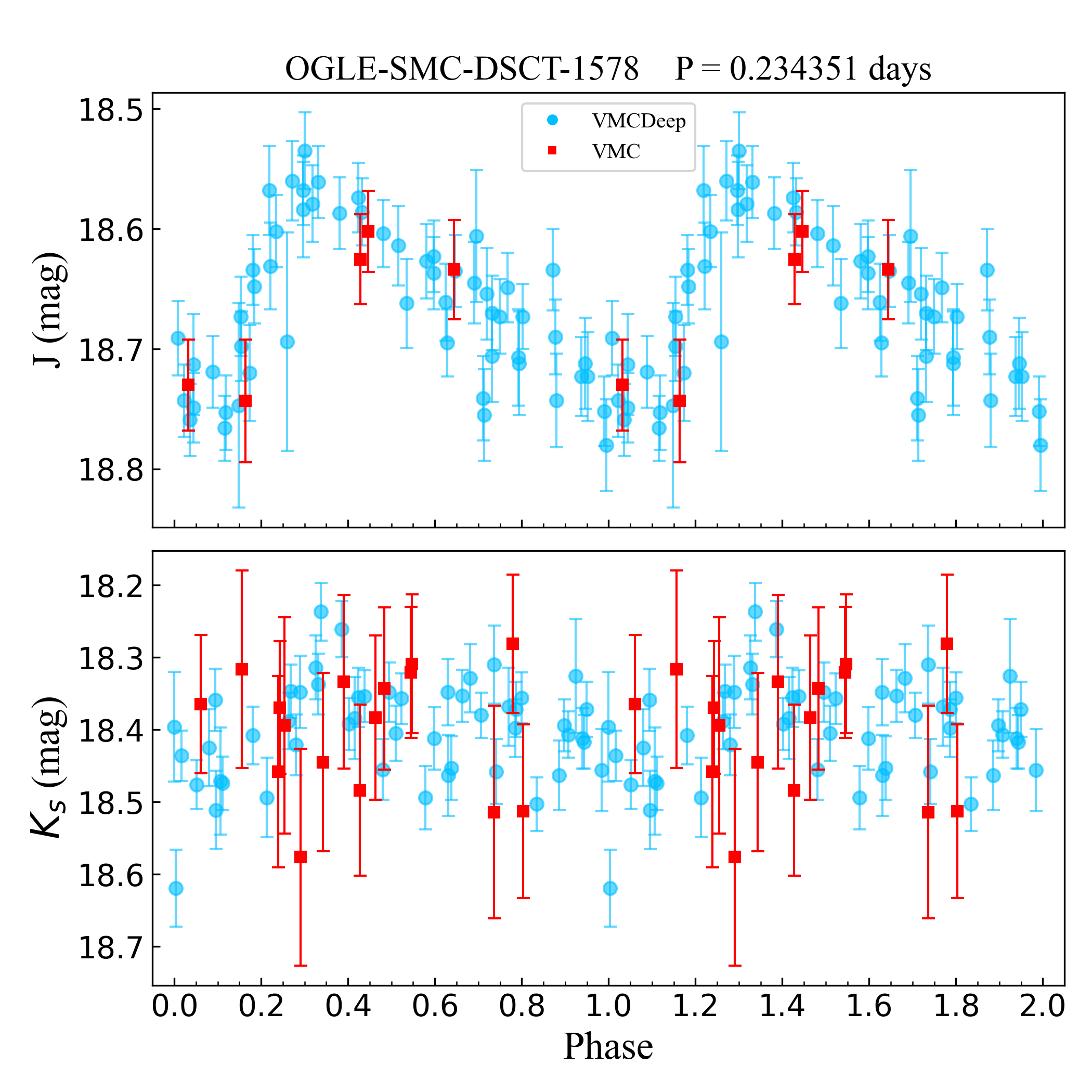}
     \caption{As Fig.\,\ref{lightcurve} but for a $\delta$ Sct star.}
        \label{lccomp}
\end{figure}

\begin{figure}
    \centering
    \includegraphics[width=8cm]{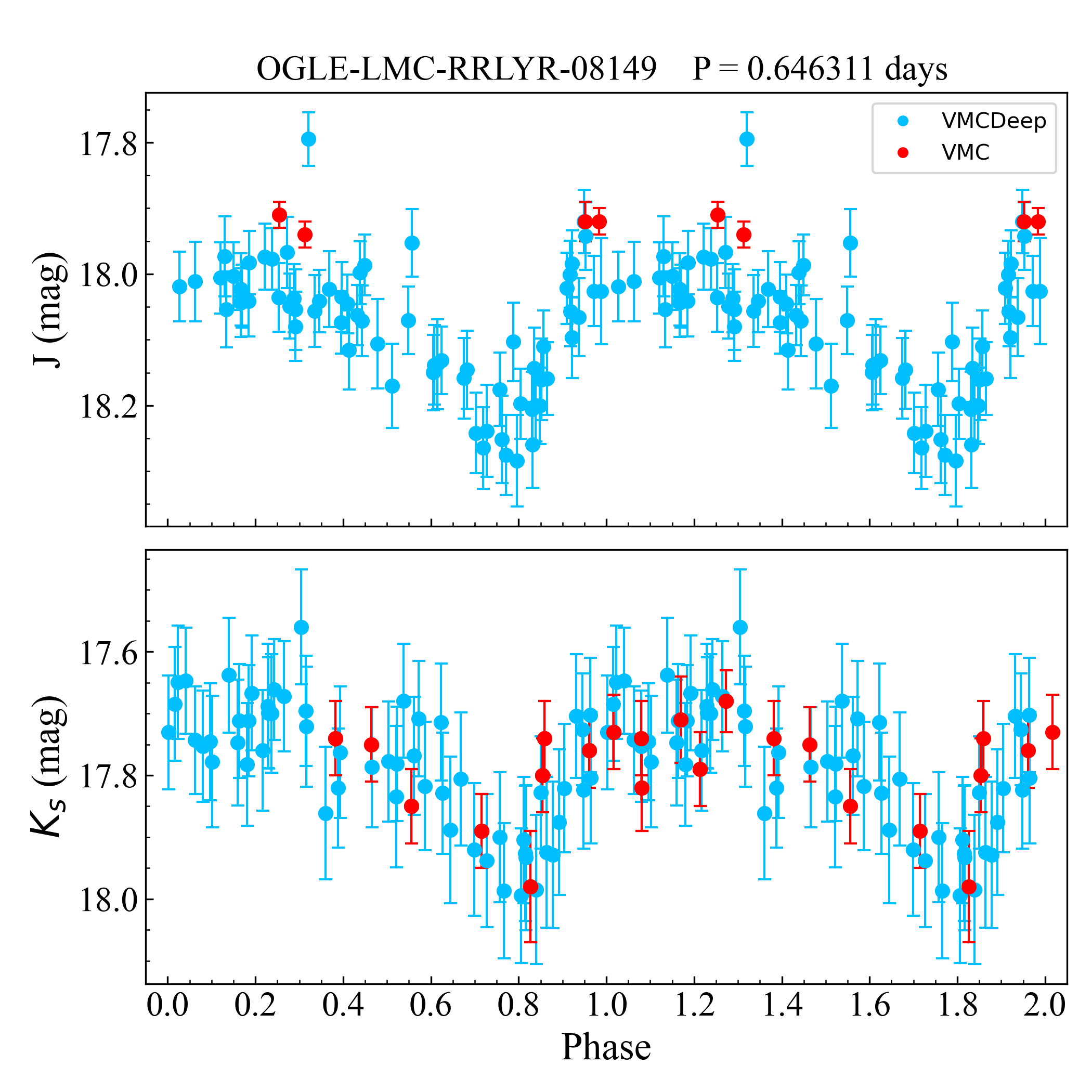}
     \caption{As Fig.\,\ref{lightcurve} but for an RR Lyrae star.}
        \label{rrl}
\end{figure}

The VMCDeep observations will allow to build well sampled $J$ and $K_\mathrm{s}$ light-curves for about 4000 and 500 RR Lyrae stars (M=0.55--0.8 M$_\odot$, age $>10$ Gyr) in the LMC and SMC, respectively, and accurately measure their average magnitudes. The magnitude variation of $\sim$1800 $\delta$ Sct (relatively young single stars) and SX Phoenicis (resulting from the merging of old stars) intermediate-mass (1.2--2.6 M$_\odot$) stars will also be sampled, thus enabling to build for the first time near-infrared period-luminosity relations for these classes of pulsating stars in the Clouds and use them as an additional standard candle. Along with their literature optical light curves they will be modelled with updated pulsation models to derive individual distances and intrinsic stellar properties (e.g.\,\citealp{mcnamara2007}). The incidence of $\delta$ Sct stars in metal-poor environments like the SMC and LMC, a quantity with a significant impact on studies of field star and star cluster evolution \citep{2016ApJ...832L..14S}, will also be determined. Moreover, RR Lyrae, $\delta$ Sct and SX Phoenicis stars will be used to measure the 3D structure of the central regions of the Clouds. These stars probe different epochs of star formation at a location where about 10\% of the OGLE data are blended. Figures \ref{lightcurve} and \ref{rrl} show the light-curve examples of Classical Cepheids and RR Lyrae stars differentiating between detections from the VMCDeep and VMC survey programmes. Figure \ref{lccomp} shows that VMCDeep provides well-sampled light-curves also for the low-amplitude and faint $\delta$ Sct stars. Note that in these figures the VMCDeep values refer to PSF photometry while the VMC values refer to VDFS photometry, performed on individual epochs. Preliminary results from the analysis of the entire sample of pulsating stars within the central regions of the Clouds show a small dispersion around the respective period-luminosity relations.

\subsection{Star formation}
\label{starformation}

\begin{figure}
    \centering
    \includegraphics[width=8cm]{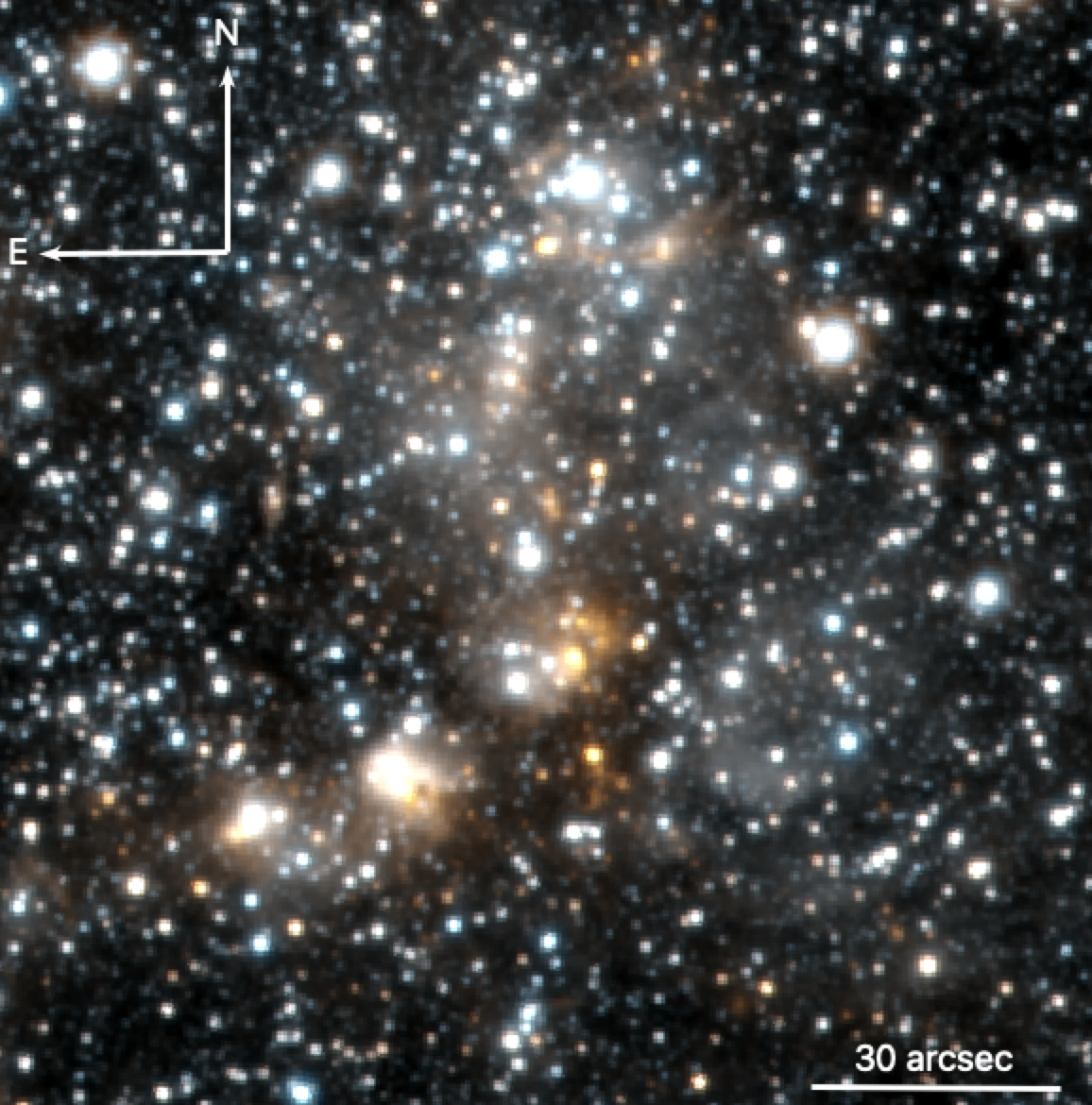} 
    \caption{Near-infrared two-colour image of the star forming region N113 in the LMC as observed in this programme. The $J$ band is shown in blue and the $K_\mathrm{s}$-band in red.}
        \label{sfregion}
\end{figure}

The central areas of the Clouds include several star forming regions (e.g.\,N113, N119, and N120 in the LMC and N19, N22, and N36 in the SMC) and massive young stellar objects (YSOs), e.g.\,over 300 in the LMC (\citealp{2012A&A...542A..66C}, \citealp{2013ApJ...778...15S}). The more homogeneous coverage, higher spatial resolution and increased exposure time (by a factor of $\sim$4) in the VMCDeep programme than for the VMC survey, will allow to study the star formation activity on large spatial scales while sampling deeper down the mass function (1.4 M$_\odot$ for ages $\le$3 Myr). Using the automated algorithm from \cite{2018A&A...620A.143Z} we will identify and study low-mass pre-main sequence stars to a similar depth (in mass) as for the outer regions sampled in VMC. Combined with the distribution of upper-main sequence stars \citep{2018ApJ...858...31S} it will provide a comprehensive picture of how star formation was spatially organised. Comparing the morphology and properties of the young populations to the underlying gas and dust distributions will also constrain the star formation rate. Furthermore, multi-epoch observations will improve the analysis of young variables;  about 400 counterparts to massive YSOs (detected in the mid-infrared) are expected, more than doubling the current sample, allowing statistically meaningful comparisons with Galactic samples (e.g.\,\citealp{2018A&A...619A..41T}, \citealp{2020MNRAS.494..458Z}). Figure \ref{sfregion} shows the star forming region N113 in the LMC whereas full-tile images are shown in Figs.\,\ref{lmcfig} and \ref{smcfig}. 

\subsection{Star formation history}
\label{sfh}

The depth and sensitivity of the VMC data have allowed a significant improvement in our knowledge of the space resolved SFH of the Clouds \citep{2018MNRAS.478.5017R,2021MNRAS.tmp.2166M}, but in the central regions where the stellar density is high crowding was a major problem. In the LMC central region, the completeness of the source detections corresponded to 75\% at $K_\mathrm{s}$<18.5 mag (cfg.\,their Fig.\,12), which means that crowding already influenced stellar populations of $\sim$1.6 Gyr old. In this case, the star formation rate is driven by the RGB and red clump regions rather than the main sequence turn-off and subgiant branch  \citep{2021MNRAS.tmp.2166M}. A similar situation occurred in the SMC, since the completeness in the central regions was significantly worse than in the other regions \citep{2018MNRAS.478.5017R}.

\begin{figure}
    \centering
    \includegraphics[width=8cm]{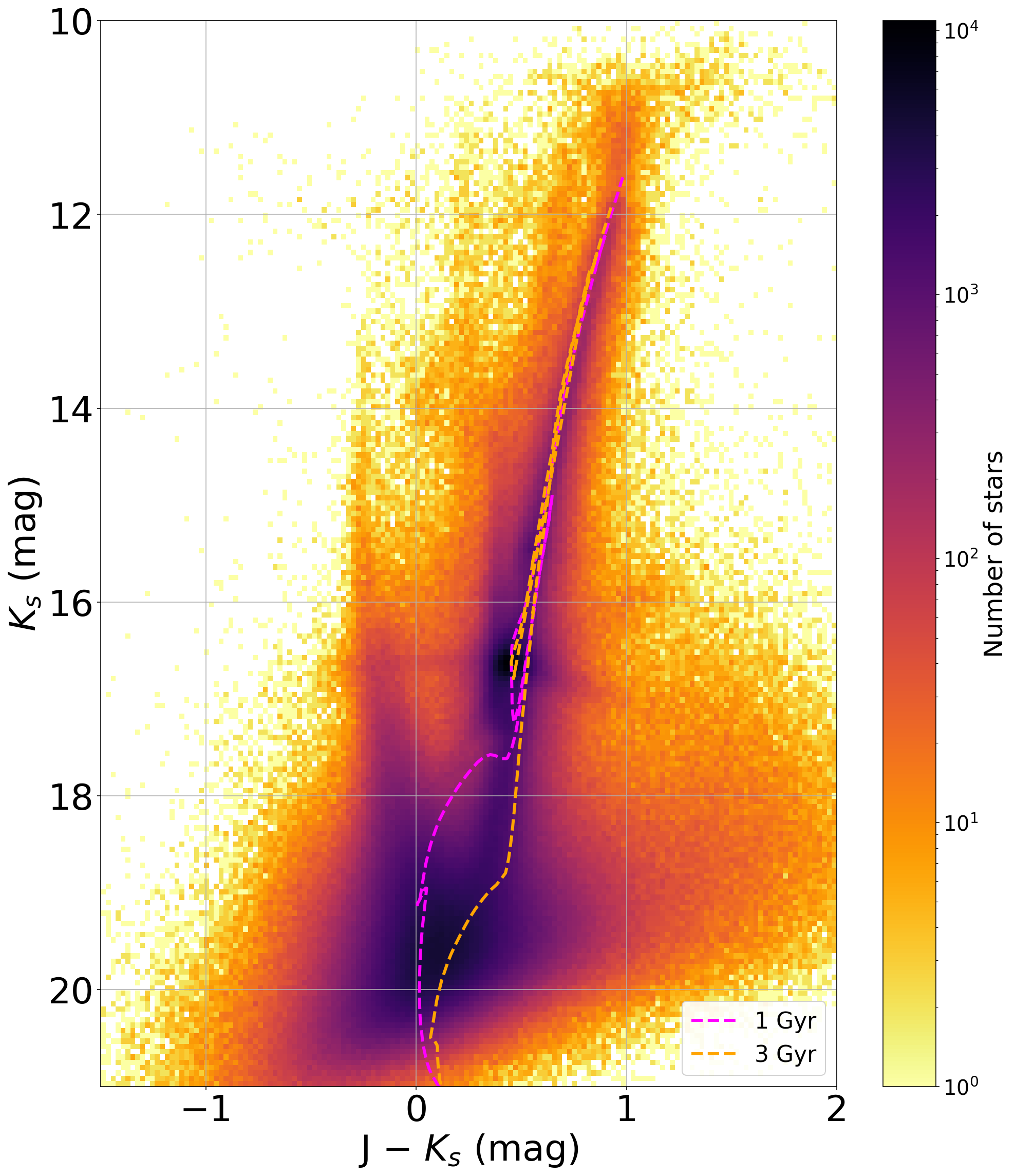} 
    \caption{Colour-magnitude diagram of the LMC from re-modelled PSF photometry of the deep and stacked observations. The PARSEC-COLIBRI isochrones refer to a distance modulus of 18.4 mag, E(B-V)=0.03 mag, [Fe/H]=--0.7 dex and --1.3 dex for ages of 1 and 3 Gyr, respectively.}
        \label{fig:sfh}
\end{figure}

The VMCDeep observations will improve the completeness of the source detections in the central regions of the Clouds and bring them in line with that in the other more external regions probed by the VMC survey. Preliminary results show that at the turn-off of the main sequence both the photometric uncertainties and sharpness distributions are significantly reduced, which will allow us to estimate a robust SFH for the intermediate-age populations ($\sim$4 Gyr old) also in the central regions. Figure \ref{fig:sfh} shows the CMD of the sources in the LMC tile with superimposed PARSEC v1.2S \citep{bressan2012} and COLIBRI \citep{marigo2013} isochrones.  In this case, the PSF for the deep and stacked observations has been re-modelled, following a different process from the one by \cite{rubele2015}, to better define and trace the CMD features. The re-modelled PSF is obtained through an iterative process designed to refine the sample of stars used for the PSF estimation. Instead of a single pass, the algorithm performs multiple iterations removing problematic stars and rebuilding the PSF model at each step, until no more stars are rejected. This leads to a cleaner, more stable PSF model, significantly improving the precision of the photometry.
 
\subsection{Red clump substructure}
\label{redclump}
Low-mass core-helium burning metal-rich stars populate the red-clump feature in the CMD, which has been extensively used to trace the line-of-sight distance and dust content of galaxies (e.g.\,\citealp{girardi2016}). In particular, in the outskirts of the SMC a bi-modal distribution has been attributed to distinct stellar populations $\sim$10 kpc apart. The foreground component, possibly of tidal origin, has been traced between about 2.5 and 10 kpc from the centre of the galaxy (e.g.\,\citealp{nidever2013,omkumar2021,elyoussoufi2021}). 

\begin{figure}
    \centering
    \includegraphics[width=8cm]{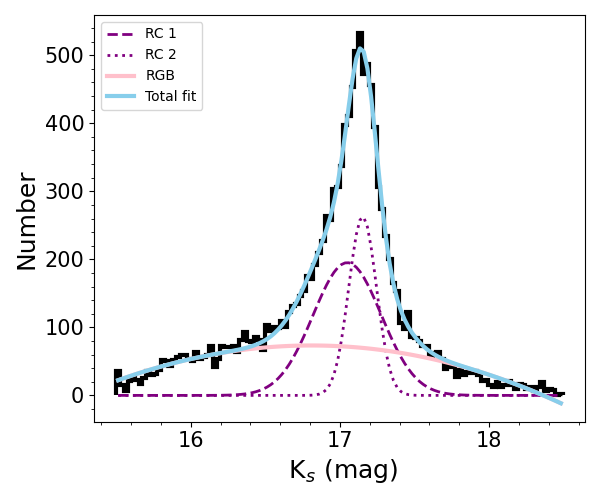} 
        \caption{Gaussian fits to the magnitude distribution of giant stars (black histogram), excluding Milky Way stars, in a region of $\sim0.16$ deg$^2$ east of the SMC centre. The best-fit total, RGB and two red-clump (RC) populations are shown with lines of different type and colour.}
        \label{fig:rc}
\end{figure}

A preliminary study based on the VMCDeep data, which is scrutinising the inner region, suggests that there is a bi-modality in the red clump distribution across the entire central 1.5 deg$^2$. Figure \ref{fig:rc} presents an example of the histogram of the stellar distribution, obtained after subtracting stars that have a high probability of belonging to the Milky Way based on the {\it Gaia} astrometry (e.g.\,\citealp{luri2021}), and the best-fits gaussians. The underlying RGB stars are best represented by a quadratic fit, whereas the dual red clump stars show two peaks that are significantly different (0.13$\pm$0.01 mag) and appear better defined than in the previous study based on the VMC data \citep{subramanian2017}. By using two rather than one gaussian to represent the red clump, the quality of the fit increases by 64\%.

\section{Conclusions}
\label{summary}
In this paper, we have described a deep VISTA observational programme designed to probe the stellar content and kinematics of the central $\sim$1.77 deg$^2$ of the Clouds. Observations were collected during the last three years of operation of the near-infrared camera in the $J$ and $K_\mathrm{s}$ bands during excellent sky conditions. The data processed through the VDFS have produced a catalogue of about 2.5 million sources with about 40 epochs per filter. Combined to the VMC survey, which covered the same regions with a mosaic of two tiles per galaxy, the deep observations represent the most spatially homogeneous observations of the Clouds' centres in the near-infrared to date. They have a high potential for studies of the stellar kinematics, the variable star populations and the structure of the galaxies as well as of star formation and the characterisation of young (variable) stellar objects. Detailed comprehensive analyses of these specific topics are ongoing and will be presented in separate papers. This also includes source catalogues with PSF photometry, which further improve source detection, and will allow us to derive more precise SFHs for the central (crowded) regions.

The data presented in this study provide useful counterparts to observations in the central regions of the Clouds at other wavelengths, for example in the X-ray domain by the extended ROentgen Survey with an Imaging Telescope Array (eROSITA; see \citealp{merloni2024} for the first data release) or in the visual domain by {\it Gaia} and Rubin. The upcoming {\it Gaia} DR4 will contain the first products from dedicated observations in regions of high stellar density including the central regions of the Clouds (Fig~\ref{map}; see \citealp{weingrill2023} for the increase of sources through the analysis of Sky Mapper images). The Rubin data will consist of hundreds of multi-band visits and sources detected from these images, as well as from difference images that add detections in dense stellar regions like the centres of the Clouds (see \citealp{acero-cuellar2026}). Ongoing observations with the {\it Euclid} space telescope will cover the SMC centre (see \citealp{mellier2025} for the coverage of the Euclid Wide Survey) before the {\it Roman} Space Telescope\footnote{https://science.nasa.gov/mission/roman-space-telescope/} begins operations.  Several sources will be spectroscopic targets for new multi-fibre spectrographs such as Multi-Object Optical and Near-infrared Spectrograph (MOONS) and 4-metre Multi Object Spectroscopic Telescope (4MOST) which both envisage dedicated observations of the central square degrees of the galaxies (see \citealp{gonzalez2020} and \citealp{cioni2019} for the respective survey strategies). In this context, studies of the central regions of the Clouds are extremely promising.

\begin{acknowledgements}
This programme is based on observations made with VISTA at the La Silla Paranal Observatory under programme IDs 105.206M, 106.201G, 108.221P and 109.23G7. 
We are grateful to C. Bell for producing one figure and to both C. Gonzalez Fernandez and Y. Kupcu Yoldas for processing the first images. We also acknowledge the assistance of Amrit Sedain in the large-scale production of the PSF photometry catalogues. We thank the Cambridge Astronomy Survey Unit (CASU) and the Wide Field Astronomy Unit (WFAU) in Edinburgh for providing calibrated data products under the support of the Science and Technology Facility Council (STFC). 
\end{acknowledgements}

\bibliographystyle{aa}

\bibliography{deep.bib}

\begin{appendix}

\section{Additional (low-quality) observations}

During the observing period additional observations were obtained that did not meet the required sky conditions. This data, except for a few observations that were interrupted due to bad weather or technical problems for which not all offset positions could be obtained, are of comparable quality to the data obtained for the VMC survey in the outer regions of Clouds. Table \ref{table:data_bad} lists their average quality control parameters. These observations may be useful to address specific scientific questions, for example, the light-curve analysis of variable stars or the determination of proper motions which benefit from extra epochs for relatively bright sources.

	\begin{table*}
		\caption{Quality control parameters for low-quality observations.}                        
		\label{table:data_bad}      
		\small
		\begin{tabular}{lcccccccccc}
			\hline \hline
			Galaxy & $\alpha$ & $\delta$ & Filter & Half tile & N & Airmass & FWHM & Ellipticity & Zero-point & Sensitivity \\
			 & (h:m:s) & ($^\circ$:$^\prime$:$^{\prime\prime}$) & & & & & ($^{\prime\prime}$) & & (mag) & (mag) \\
			\hline
            SMC & 00:50:16.06 & --73:10:40.92 & $J$ & LFT & 16 & 1.59$\pm$0.06 & 1.05$\pm$0.29 & 0.09$\pm$0.05 & 23.80$\pm$0.16 & 19.63$\pm$0.28 \\
             & & & $J$ & RHT & 23 & 1.60$\pm$0.05 & 1.01$\pm$0.12 & 0.10$\pm$0.06 & 23.83$\pm$0.03 & 19.71$\pm$0.13 \\
             & & & $K_\mathrm{s}$ & LFT & 11 & 1.54$\pm$0.04 & 0.92$\pm$0.16 & 0.08$\pm$0.03 & 23.08$\pm$0.01 & 18.80$\pm$0.13 \\
             & & & $K_\mathrm{s}$ & RHT & 11 & 1.55$\pm$0.05 & 0.92$\pm$0.14 & 0.08$\pm$0.02 & 23.07$\pm$0.01 & 18.76$\pm$0.11 \\
            LMC & 05:20:54.47 & --69:34:43.32 & $J$ & LFT & 35 & 1.51$\pm$0.08 & 0.90$\pm$0.11 & 0.10$\pm$0.04 & 23.65$\pm$0.55 & 19.01$\pm$0.46 \\
             & & & $J$ & RHT & 39 & 1.51$\pm$0.08 & 0.91$\pm$0.11 & 0.09$\pm$0.04 & 23.74$\pm$0.19 & 19.15$\pm$0.16  \\
             & & & $K_\mathrm{s}$ & LFT & 26 & 1.46$\pm$0.07 & 0.88$\pm$0.08 & 0.06$\pm$0.02 & 23.02$\pm$0.13 & 18.37$\pm$0.15 \\
             & & & $K_\mathrm{s}$ & RHT & 25 & 1.47$\pm$0.06 & 0.94$\pm$0.16 & 0.08$\pm$0.02 & 23.04$\pm$0.03 & 18.38$\pm$0.13 \\
			\hline
		\end{tabular}
		\tablefoot{The coordinates are those of the tile centres. LFT corresponds to the tile pattern {\it Tile3px} and RHT to {\it Tile3nx}. N is the number of epochs. The sensitivity corresponds to 5$\sigma$ point source limiting magnitudes.}
	\end{table*}

\section{Photometric shifts}
\label{photshifts} 

\begin{figure}
\centering
\includegraphics[width=\hsize]{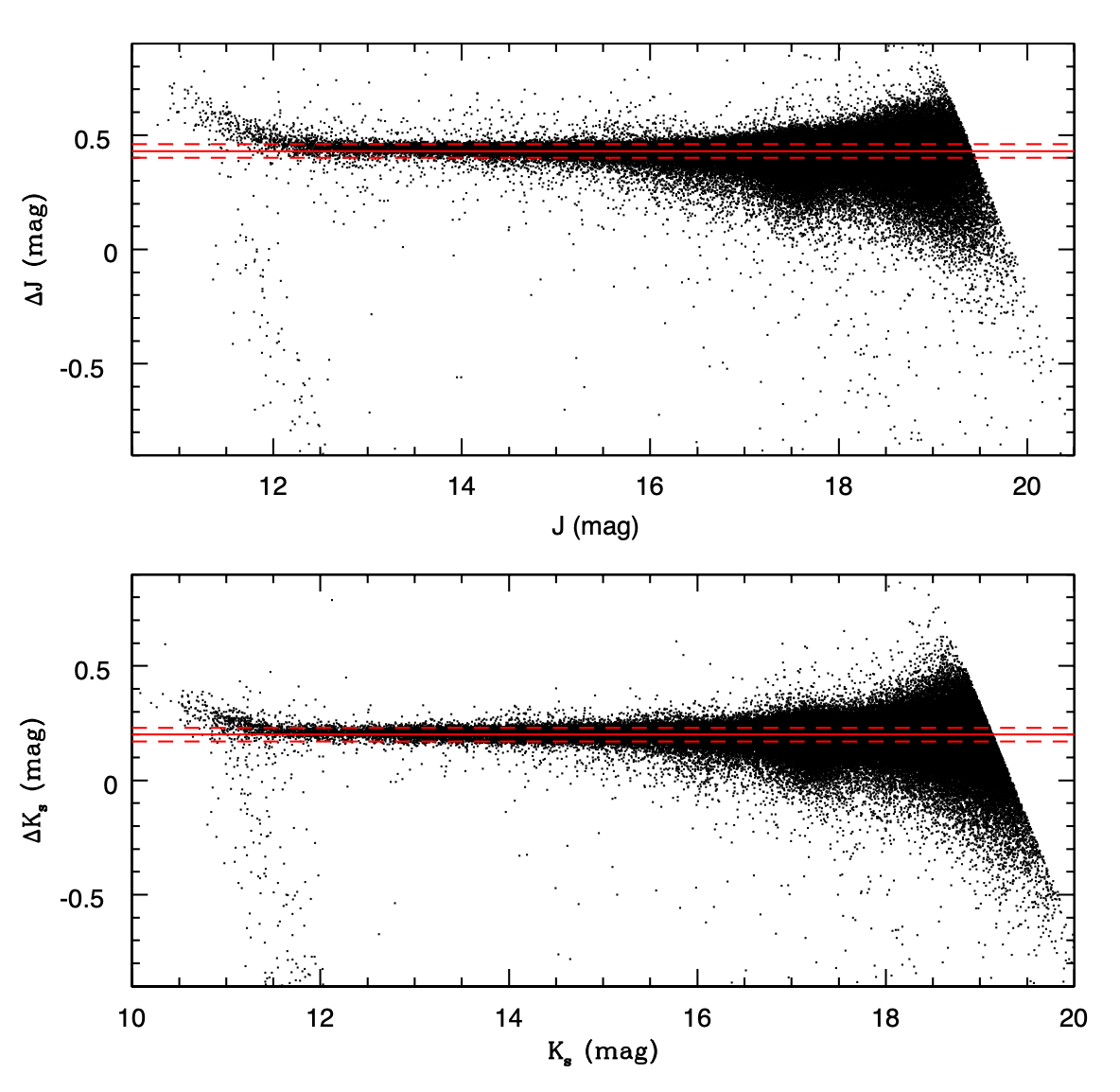}
\caption{Photometric shifts between VMCDeep and VMC sources in the SMC for calibrating the PSF photometry.}
\label{fig:shifts}
\end{figure}

\begin{figure*}
\centering
\includegraphics[width=8.9cm]{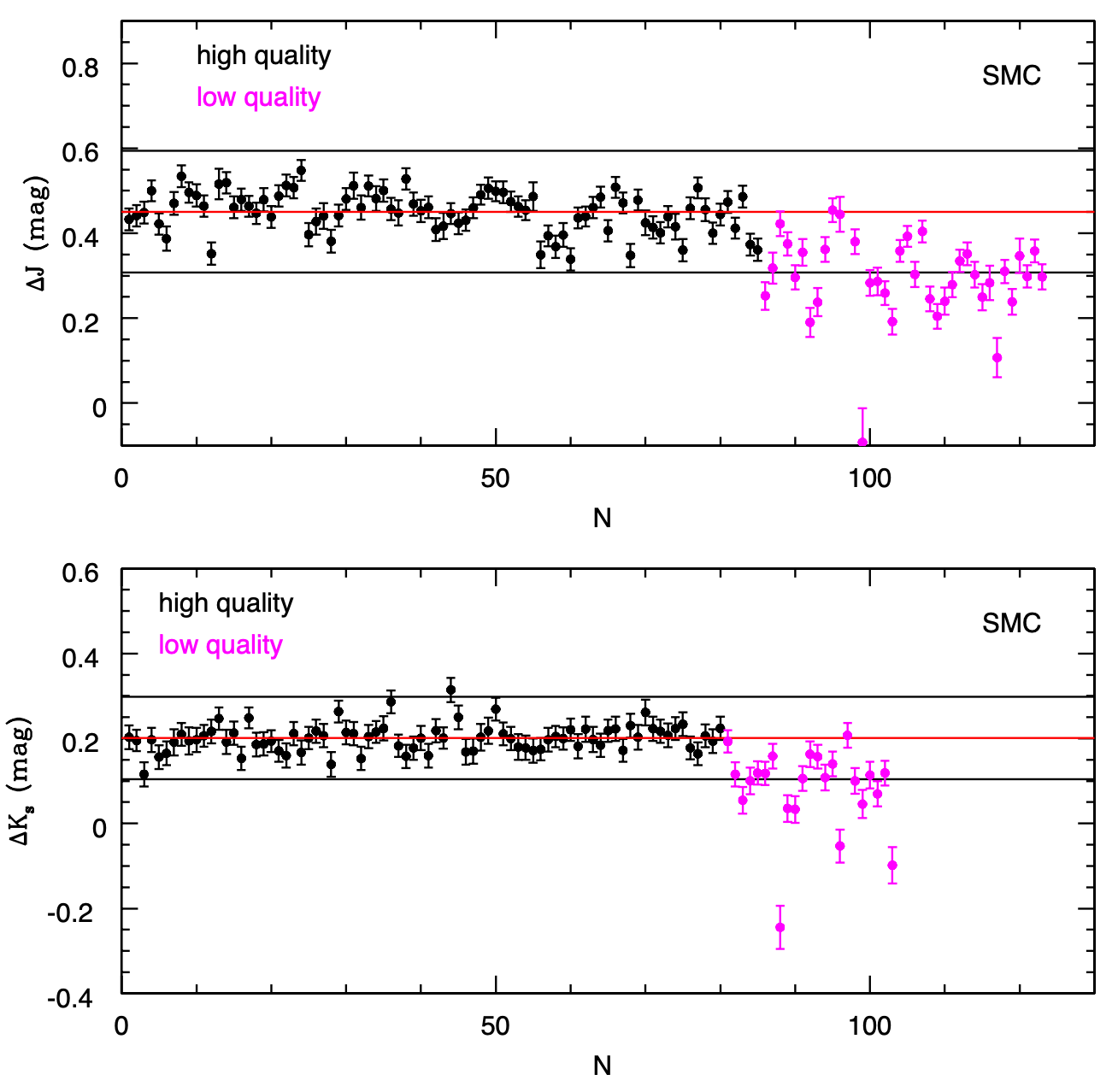}
\includegraphics[width=9cm]{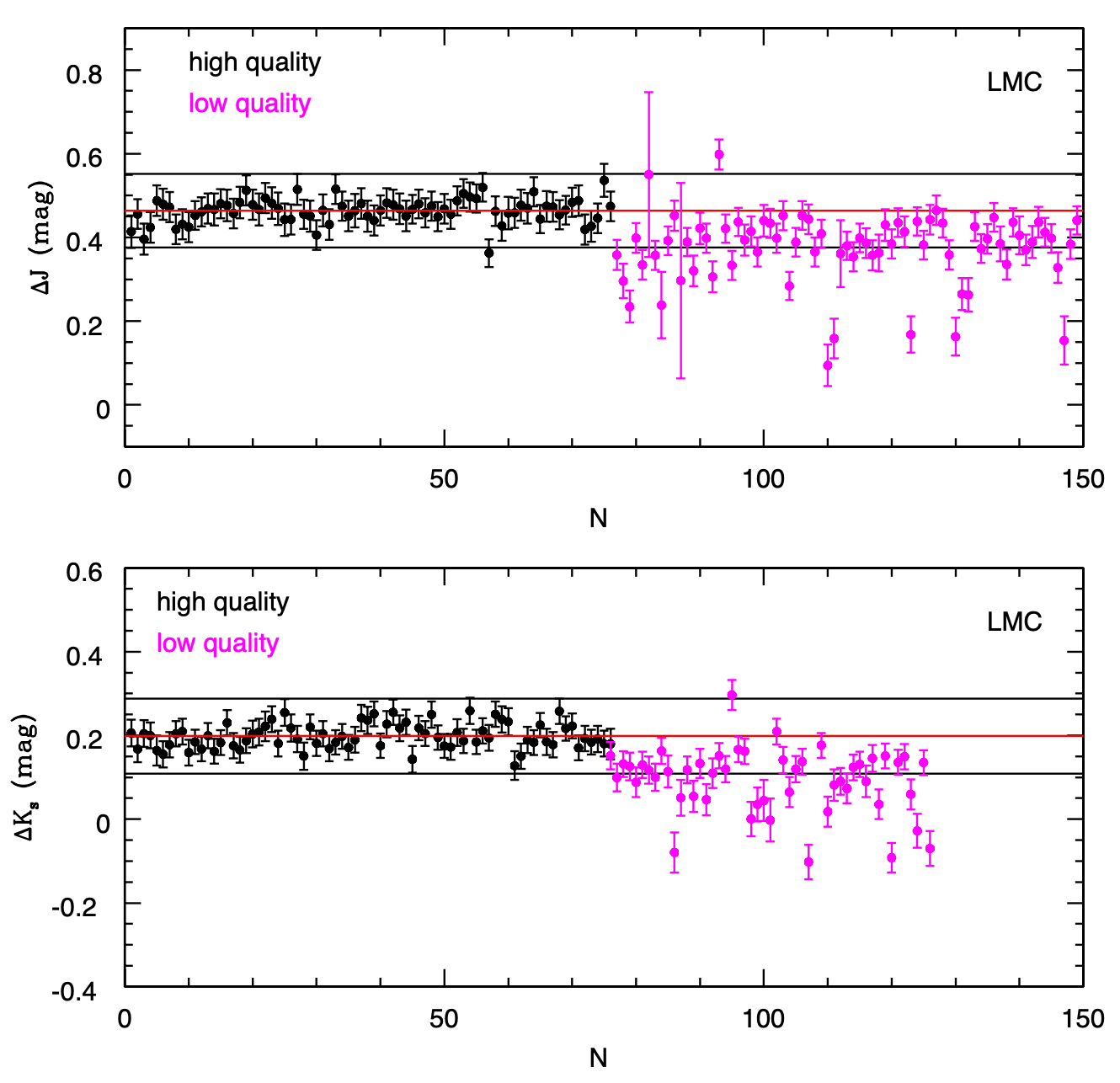}
\caption{Photometric shifts for single half-tile images between VMCDeep and VMC sources in the SMC (left) and LMC (right)  for calibrating the PSF photometry. The horizontal axis is a sequential number (N) with high-quality observation first (black) and low-quality observations last (magenta). Horizontal lines show the average of the shifts obtained only from the high-quality observations with the 1-$\sigma$ range. }
\label{singleshiftsfig}
\end{figure*}

The PSF photometry is adjusted to the photometry from the VDFS to place it on an absolute scale. Table \ref{shifts} shows the median shifts in the $J$ and $K_\mathrm{s}$ filters for the LMC and SMC sources. These are calculated using unique VDFS detections in both filters, with only minor quality issues ({\it ppErrBits}<256) and with photometric uncertainties <0.1 mag. Figure \ref{fig:shifts} shows as an example, the SMC sources and shifts. Note that at the bright end the stellar population is dominated by asymptotic giant branch stars which are pulsating variables and cause the curvature towards positive values. On the contrary, sources influenced by saturation produce largely deviating negative values. Individual shifts for each  half-tile image are given in Tab.\,\ref{singleshiftstab} where each line refers to one image and indicates: the name of the catalogue, the corresponding shift and its uncertainty. The name includes the type of galaxy, filter, type of half tile and observing date as well as the number of the first images in the set of offsets that produce the half tile. The filter is followed by a number which identifies the planned epoch of observation. Low-quality observations result in photometric shifts that are systematically below the average of the shifts derived for the high-quality observations, with several measurements outside the 1-$\sigma$ range and larger individual uncertainties. The individual shifts are also shown in Fig.\,\ref{singleshiftsfig} for both high- and low-quality observations. Table \ref{singleshiftstab} is only available at CDS.

Table \ref{shifts} also includes photometric shifts between the PSF catalogues from the VMCDeep and VMC survey programmes. These are computed utilising all sources in common. They are consistent within the uncertainties with the calibration of the VMCDeep PSF catalogue above with respect to the VDFS photometry.

\begin{table}
\setlength{\tabcolsep}{3pt}
	\caption{Median photometric shifts.}
	\label{shifts}
	\small
	\centering
	\begin{tabular}{lcccc}
	\hline\hline
	Galaxy & \multicolumn{2}{c}{VDFS$_\mathrm{VMCDeep}-$PSF$_\mathrm{VMCDeep}$} & \multicolumn{2}{c}{PSF$_\mathrm{VMC}-$PSF$_\mathrm{VMCDeep}$} \\ 
	 & $J$ & $K_\mathrm{s}$ &  $J$ & $K_\mathrm{s}$\\
	 & (mag) & (mag) & (mag) & (mag)\\
	\hline
	LMC & 0.45$\pm$0.03 & 0.28$\pm$0.03 & 0.41$\pm$0.07 & 0.26$\pm$0.06\\
	SMC & 0.47$\pm$0.03 & 0.22$\pm$0.02 & 0.42$\pm$0.05 & 0.22$\pm$0.05\\
	\hline
	\end{tabular}
\end{table}

\begin{table}
\setlength{\tabcolsep}{3pt}
\caption{Example of individual photometric shifts.}
\label{singleshiftstab}
\small
\centering
\begin{tabular}{lcc}
\hline\hline
Name & Shift & Uncertainty \\
 & (mag) & (mag) \\
 \hline
psf\_SMC\_LFT\_J1\_20220808\_00251.cat & 0.433 &  0.025 \\
psf\_SMC\_LFT\_J2\_20220905\_00310.cat &  0.441 &  0.025 \\
psf\_SMC\_LFT\_J3\_20220915\_00262.cat &  0.449 &  0.026 \\
psf\_SMC\_LFT\_J4\_20220915\_00289.cat &  0.500 &  0.025 \\
psf\_SMC\_LFT\_J5\_20210820\_00252.cat &  0.422 &  0.025 \\
psf\_SMC\_LFT\_J5\_20220628\_01009.cat &  0.387 & 0.028 \\
psf\_SMC\_LFT\_J6\_20210820\_00333.cat &  0.470 & 0.027 \\
psf\_SMC\_LFT\_J6\_20220629\_01039.cat &  0.534 & 0.026 \\
psf\_SMC\_LFT\_J7\_20210821\_00308.cat &  0.496 & 0.024 \\
\hline
\end{tabular}
\end{table}

\section{Maps from the VMC survey}

\begin{figure}
\centering
\includegraphics[width=\hsize]{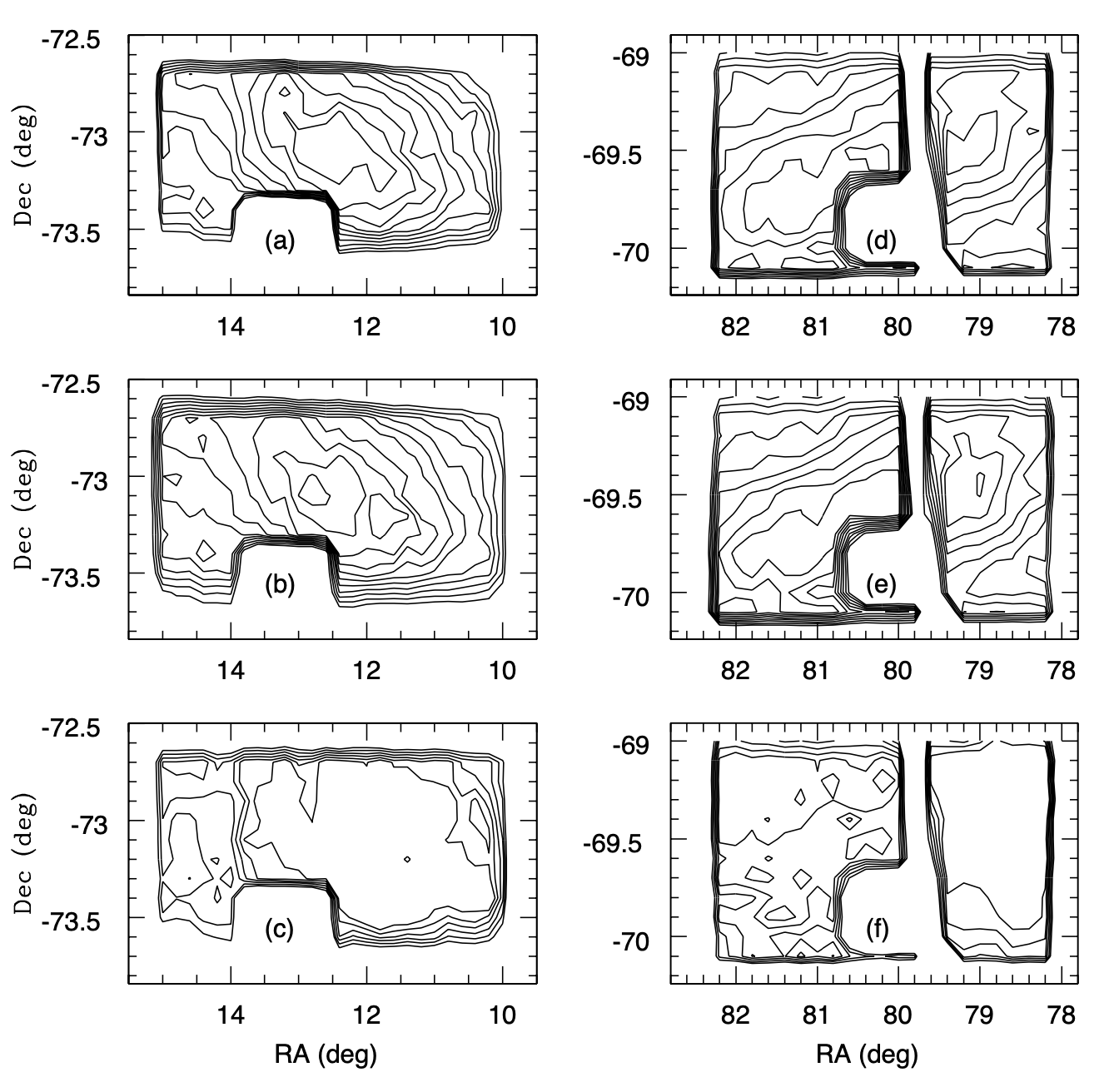}
\caption{As in Fig.~\ref{maps} but using data from the VMC survey.}
\label{vmcmaps}
\end{figure}

The VMC survey observed the Clouds by mosaicking 68 and 27 tiles in the LMC and SMC, respectively. Figure \ref{vmcmaps} shows the coverage of the central regions from this programme by the VMC observations. Both regions result from the observations of two adjacent VMC tiles. The missing rectangular areas are due to detector \#16. In the SMC this causes a lack of sources to the south of the centre whereas in the LMC, where the position angle of the tiles is rotated by +90 deg, this is very close to the centre itself. In addition, in the LMC there is a lack of sources along a strip through the centre. This corresponds to the location of the underexposed region of a tile where the exposure time is shared between the adjacent tiles. The same occurs in the SMC in declination. A selection of sources with only minor quality issues ({\it ppErrBits}<256) excludes stars observed in both detector \#16 and the ears of tiles. The tile overlaps for the other sides correspond to 30\arcsec, but due to the full depth of the observations at their location there are no obvious discontinuities in the mosaic.

\section{Confusion in the VMC survey}
\begin{figure}
    \centering
    \includegraphics[width=\hsize]{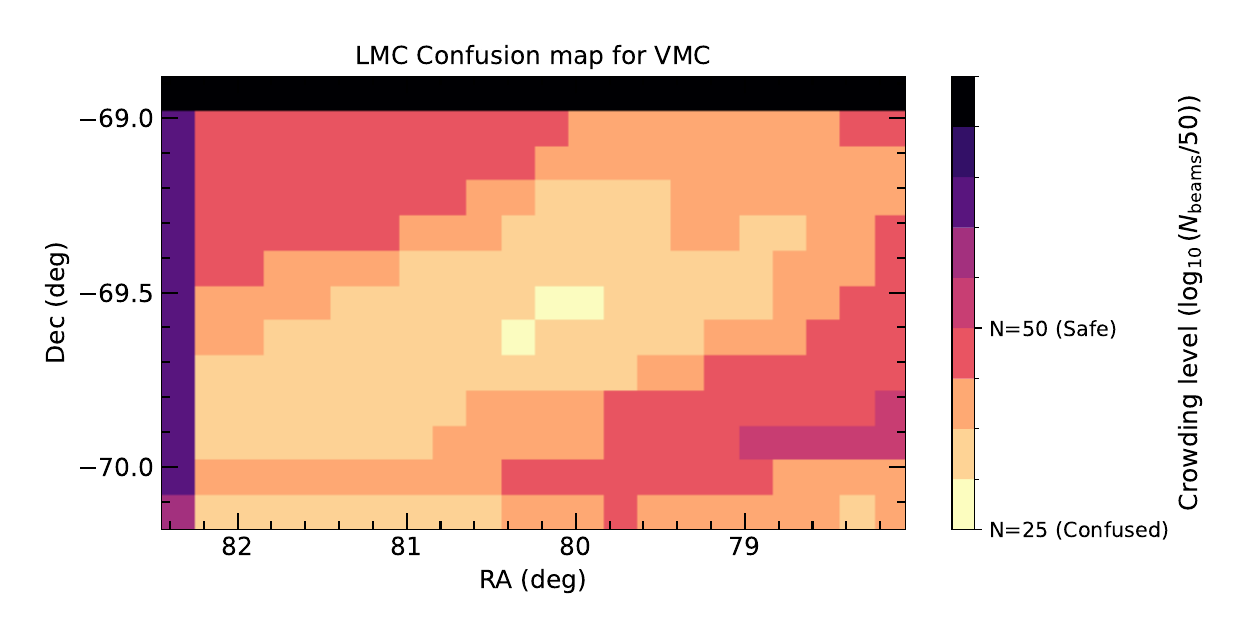}
    \includegraphics[width=\hsize]{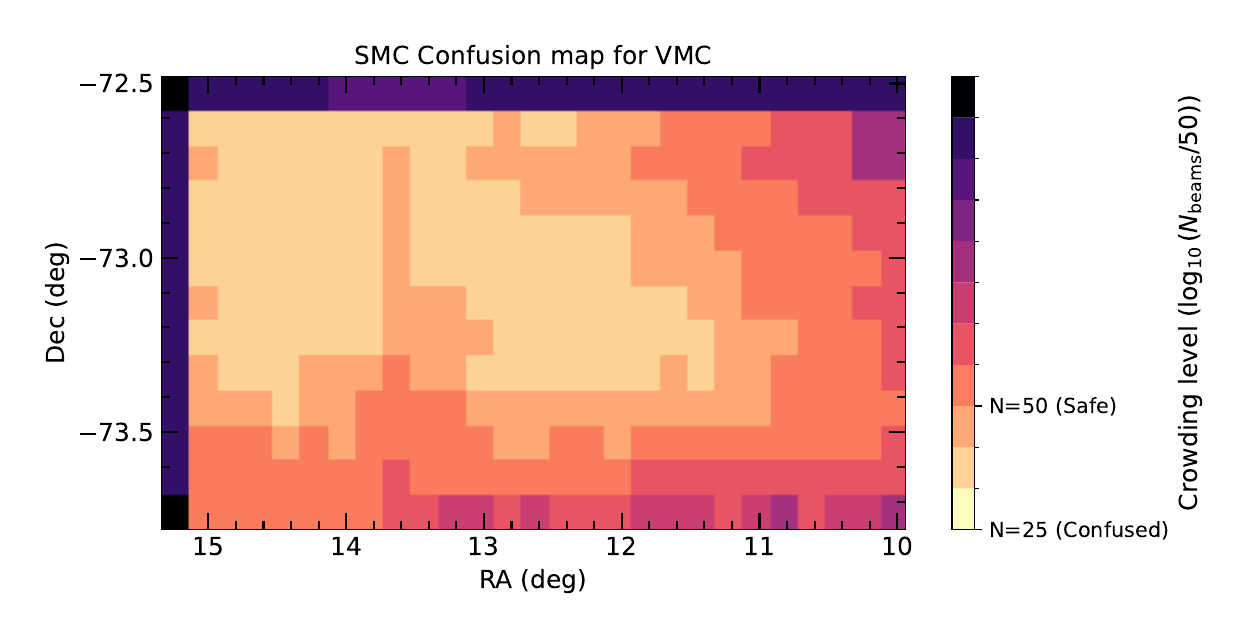}
    \caption{Confusion maps for bins of 0.2$\times$0.1 deg$^2$ in the LMC (top) and SMC (bottom) central regions from the VMC survey. The threshold number of beams (N=50) and the critical value (N=25) are indicated in the colour bar.}
        \label{confusionvmc}
\end{figure}

Confusion maps of the sources detected in the central regions of the LMC and SMC from the VMC survey are shown in Fig.\,\ref{confusionvmc}. They have been created as in Sec.\,\ref{sec:confusion} and can be compared with those from the VMCDeep observations shown in Fig.\,\ref{confusiondeep}.

\section{Tile images}

\begin{figure*}[h]
\centering
\includegraphics[width=\hsize]{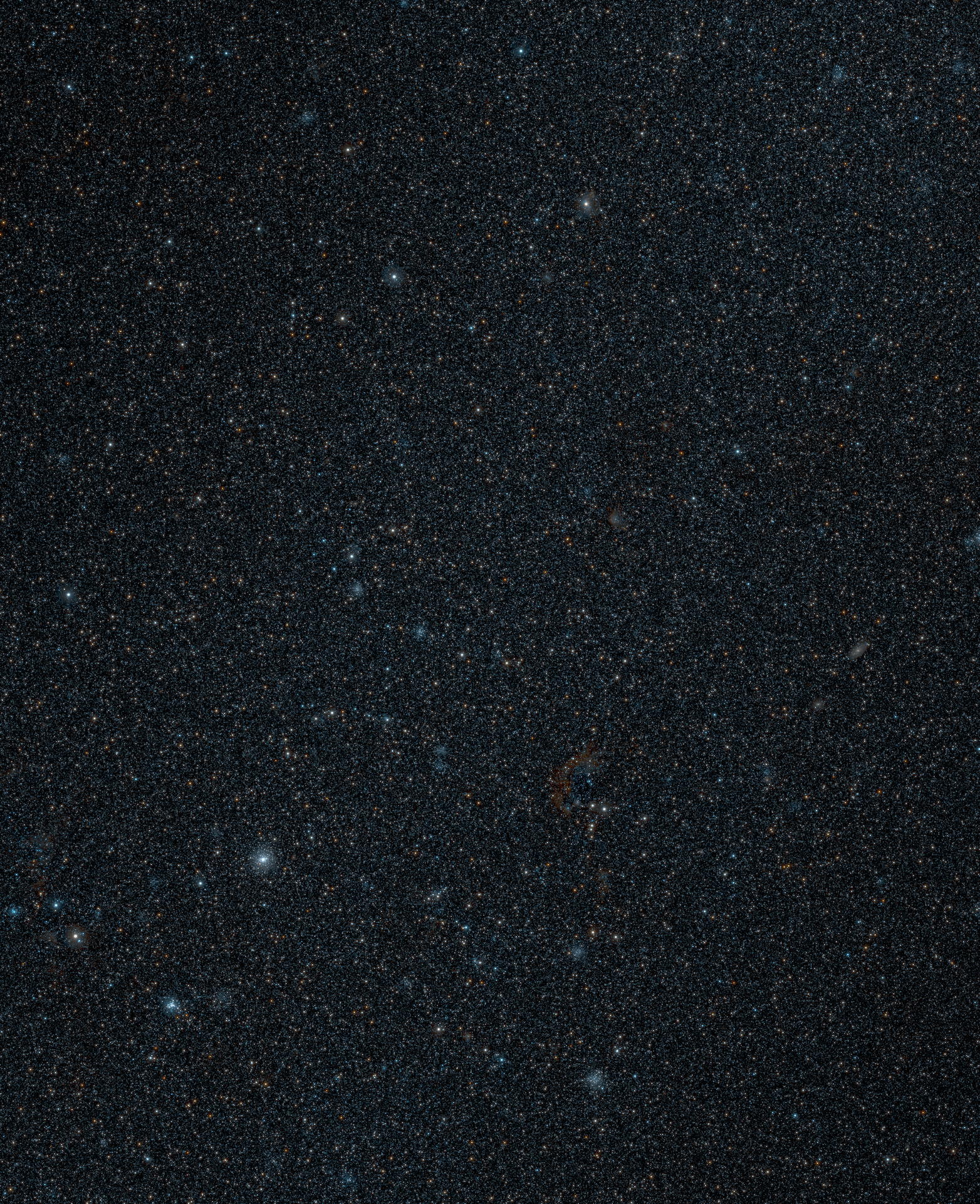}
\caption{Colour-composite image of most of the LMC tile where $J$ is in blue and $K_\mathrm{s}$ is in red. East is to the left and North at the top. Several stellar clusters are clearly visible while bright blue stars contrast a red gaseous arc near the centre of the field.}
\label{lmcfig}
\end{figure*}

\begin{figure*}[h]
\centering
\includegraphics[width=\hsize]{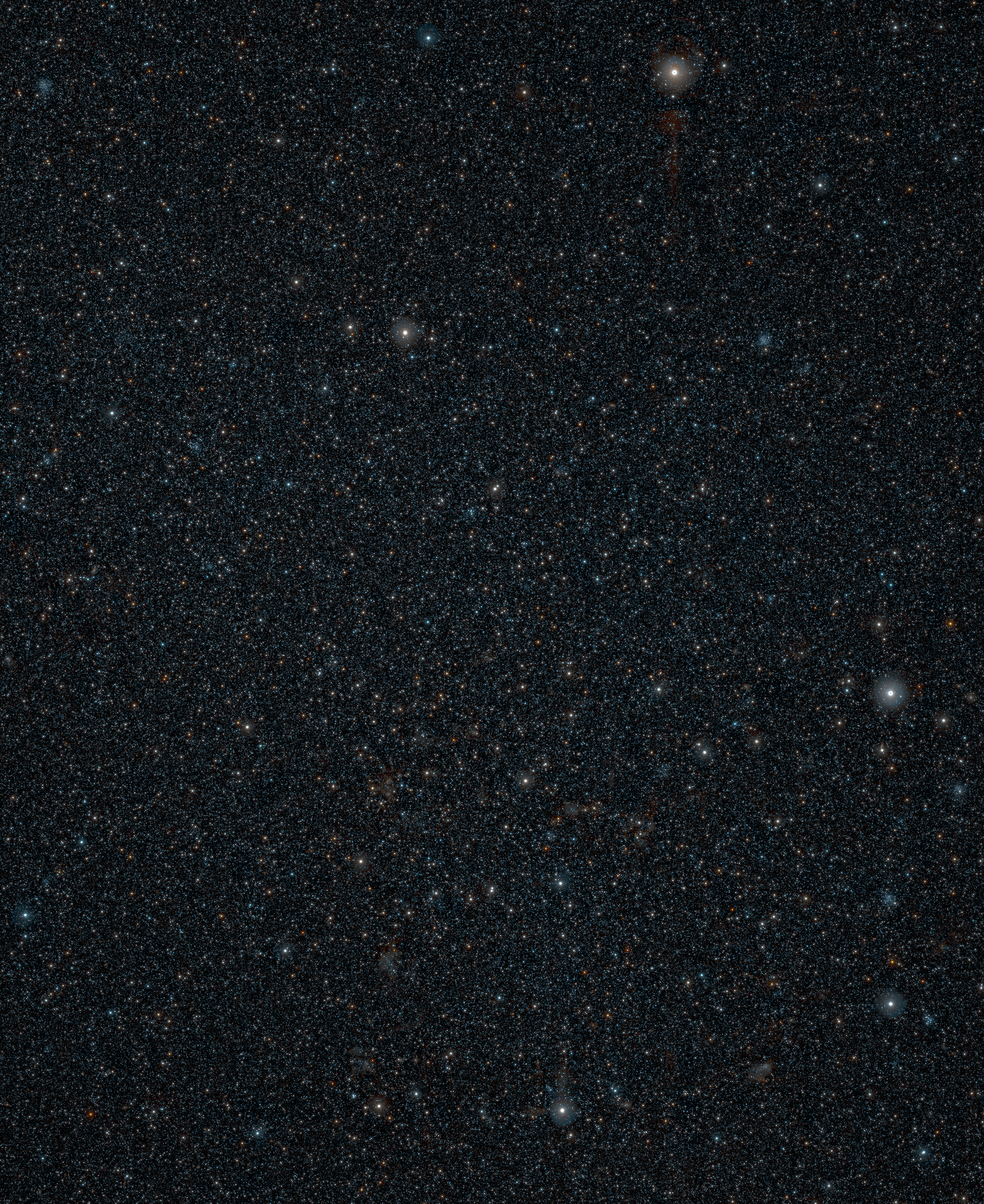}
\caption{Colour-composite image of most of the SMC tile where $J$ is in blue and $K_\mathrm{s}$ is in red. East is to the left and North at the top. Bright red stars are scattered in the midst of the density field while the glow of the brightest stars veils the neighbouring sources.}
\label{smcfig}
\end{figure*}

Figures \ref{lmcfig} and \ref{smcfig} show two-colours mosaic images of the LMC and SMC central regions from this programme. They have been obtained after combining all good-quality observations.

\end{appendix}

\end{document}